\documentclass{article}

\PassOptionsToPackage{numbers}{natbib}
\usepackage[final]{neurips_2025}

\usepackage[utf8]{inputenc} 
\usepackage[T1]{fontenc}    
\usepackage{hyperref}       
\usepackage{url}            
\usepackage{booktabs}       
\usepackage{amsfonts}       
\usepackage{nicefrac}       
\usepackage{microtype}      
\usepackage{xcolor}         

\usepackage{amsmath}
\usepackage{gensymb}
\usepackage{subcaption}
\usepackage{graphicx}
\usepackage{float}
\title{Connectome-Based Modelling Reveals Orientation Maps in the \textit{Drosophila} Optic Lobe}

\author{%
Jia-Nuo Liew$^{1,2,}$\thanks{Equal contribution.} ,
Shenghan Lin$^{3,}$\footnotemark[1] ,
Bowen Chen$^{2,4,6,7}$ , 
Wei Zhang$^{1,2}$  , \\
\textbf{Xiaowei Zhu$^{4}$  ,}
\textbf{Wei Zhang$^{2,4,6,7,}$\thanks{Corresponding author.} ,}
\textbf{Xiaolin Hu$^{1,2,8,}$\footnotemark[2]} \\[0.5em]
$^{1}$Department of Computer Science and Technology, BNRist, Tsinghua University, Beijing 100084, China \\
$^{2}$IDG/McGovern Institute of Brain Research, Tsinghua University, Beijing 100084, China \\
$^{3}$Zhili College, Tsinghua University, Beijing 100084, China \\
$^{4}$School of Life Sciences, Tsinghua University, Beijing 100084, China \\
$^{5}$Ant Group, China \\
$^{6}$State Key Laboratory of Membrane Biology, Tsinghua University, Beijing 100084, China \\
$^{7}$Tsinghua-Peking Center for Life Sciences, Tsinghua University, Beijing 100084, China \\
$^{8}$Chinese Institute for Brain Research (CIBR), Beijing 100010, China\\[0.5em]
\texttt{\{liujn24, linsh24, cbw21, zhangw23\}@mails.tsinghua.edu.cn} \\
\texttt{robert.zxw@antgroup.com} \\
\texttt{\{wei\_zhang, xlhu\}@mail.tsinghua.edu.cn} \\
}

\begin{document}


\setcounter{figure}{0}
\renewcommand{\thefigure}{E\arabic{figure}}

\begin{center}
    {\LARGE\bfseries Errata}
\end{center}

\vspace{0.5em}

This erratum clarifies the analyses presented in Figures 4 and 5a of the
original paper.

\noindent\textbf{Figure 4.}
In the original Figure 4 analysis, Dm3 and Dm15 neurons were excluded because
they form a compact, stratified band within the distal medulla and exhibit
largely type-specific orientation preferences. The purpose of this exclusion
was to make the broader, spatially continuous orientation structure of the Dm
population easier to visualise. Repeating the analysis while including Dm3 and
Dm15 neurons still reveals a coherent orientation map (Figure~\ref{fig:errataE1}).
Their inclusion introduces a prominent stratified band and increases local
type-dependent structure, but does not eliminate the mesoscale spatial
organisation of preferred orientations or the overall orientation-map pattern.

\begin{figure}[H]
    \centering
    \includegraphics[width=0.72\linewidth]
    {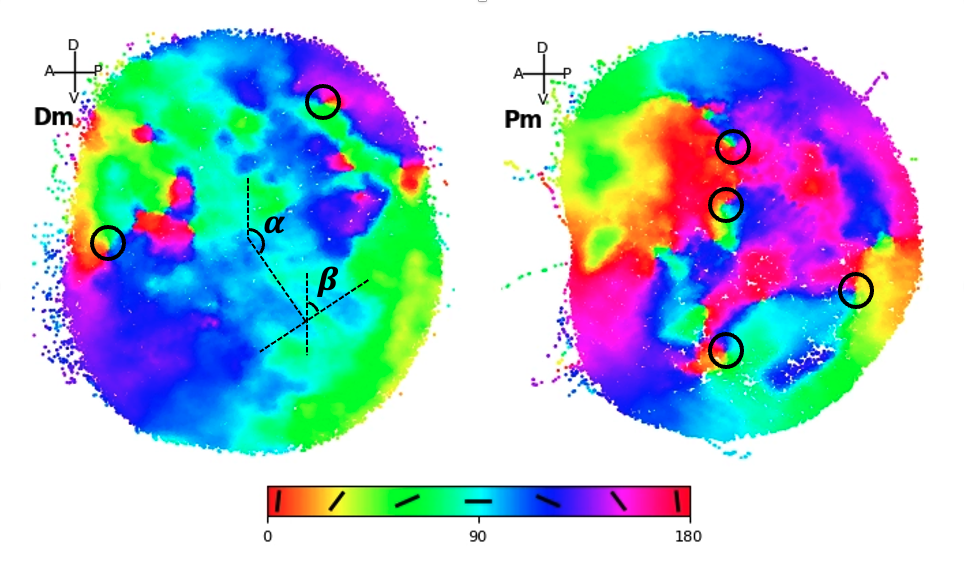}
    \caption{\textbf{Orientation maps when Dm3 and Dm15 neuron types were included.}
    Preferred-orientation maps across optic-lobe layers after spatial smoothing,
    colour-coded by preferred orientation angle. The spatially organised
    orientation-map structure remains visible when Dm3 and Dm15 neurons are
    included in the analysis.}
    \label{fig:errataE1}
\end{figure}

\vspace{-0.6em}

\noindent\textbf{Figure 5a.}
Figure 5a has been redrawn to show the Dm orientation-column examples more
clearly (Figure~\ref{fig:errataE2}). The revised figure explicitly identifies
the compact layer formed by Dm3 and Dm15 neurons and distinguishes this
type-specific stratification from the orientation organisation of the
surrounding Dm neurons.

\begin{figure}[H]
    \centering
    \includegraphics[width=0.72\linewidth]
    {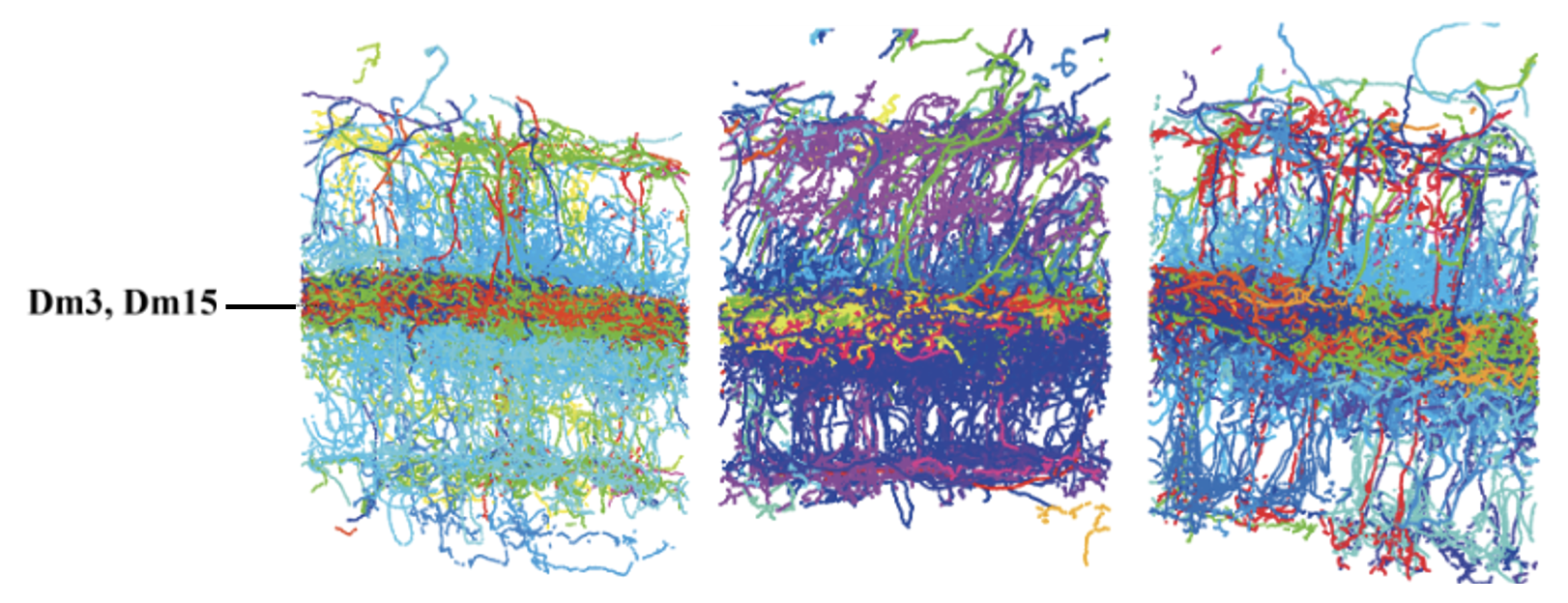}
    \caption{\textbf{Redrawn version of original Figure 5a.}
    Dm orientation-column examples showing the compact layer formed by Dm3
    and Dm15 neurons and the orientation organisation of the surrounding
    Dm neurons.}
    \label{fig:errataE2}
\end{figure}

\vspace{-0.5em}

\noindent
These clarifications do not alter the conclusions of the paper:
orientation preferences remain spatially organised within the
\textit{Drosophila} optic lobe, and orientation-map structure is observed
both with and without Dm3 and Dm15 neurons included in the analysis.


\clearpage

\setcounter{figure}{0}
\renewcommand{\thefigure}{\arabic{figure}}

\maketitle

\begin{abstract}
  The ability to extract oriented edges from visual input is a core computation across animal vision systems. Orientation maps, long associated with the layered architecture of the mammalian visual cortex, systematically organise neurons by their preferred edge orientation. Despite lacking cortical structures, the \textit{ Drosophila melanogaster} brain contains feature-selective neurons and exhibits complex visual detection capacity, raising the question of whether map-like vision representations can emerge without cortical infrastructure. We integrate a complete fruit fly brain connectome with biologically grounded spiking neuron models to simulate neuroprocessing in the fly visual system. By driving the network with oriented stimuli and analysing downstream responses, we show that coherent orientation maps can emerge from purely connectome-constrained dynamics. These results suggest that species of independent origin could evolve similar visual structures.
\end{abstract}

\section{Introduction}
The ability to extract oriented edges from visual input is a core computation across animal vision systems \cite{Hubel_Wiesel_1962}. Orientation selectivity is canonically attributed to the primary visual cortex (V1), where neurons respond selectively to specific edge orientations. \citet{Hubel_Wiesel, Hubel_Wiesel_1962} first demonstrated this modular structure in cats. The findings were later substantiated by electrophysiology, optical imaging, and detailed anatomical mapping \cite{Blasdel_Obermayer, Ohki_Chung}. These orientation-selective responses are embedded within columnar and laminar architecture and are thought to arise from a combination of spatially organised feedforward inputs and recurrent cortical dynamics \cite{Angelucci_Bressloff, crair_gillespie, Ferster_Miller, Hansel_Vreeswijk, Vita_Orsi}. 

Orientation selectivity has been directly observed in \textit{Drosophila melanogaster} visual system. In particular, T4 and T5 neurons - traditionally known for direction selectivity - also exhibit robust orientation tuning, which sharpens motion detection \cite{Fisher_Silies}. These findings provide physiological evidence that orientation selectivity exists at the neuron level. Building on this, recent studies have hypothesised that neurons in early visual stages, particularly in the medulla, could support orientation selectivity, based solely on synaptic connectivity patterns from structural data \cite{Seung}. Though the question of whether such orientation selectivity is organised into coherent maps across the visual system remains unknown. Given the compact, non-layered architecture of the \textit{Drosophila} brain and the absence of large-scale recurrent loops seen in vertebrate cortex, it is unclear whether this system can support the emergence of global tuning structures such as orientation maps \cite{Borst_Helmstaedter, Clark_Bursztyn, Maisak_Haag}. 

\begin{figure}[tp]
    \centering
    \begin{subfigure}[b]{0.4\linewidth} 
        \centering
        \includegraphics[width=\textwidth]{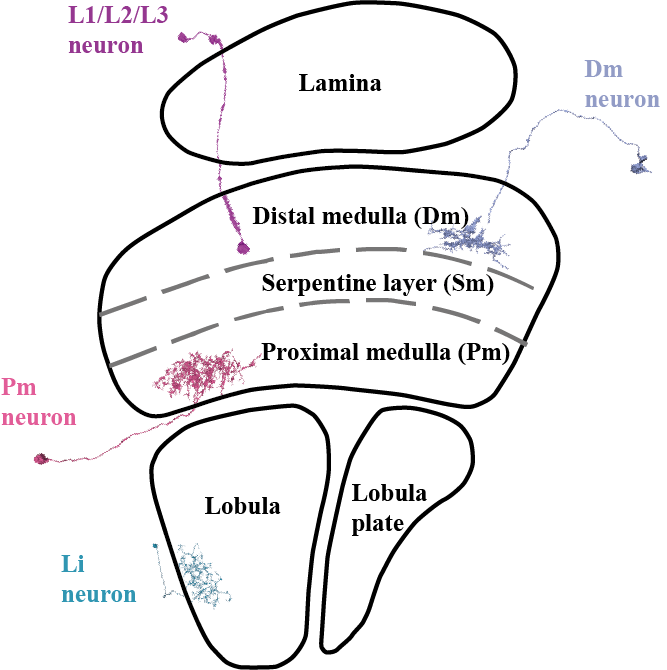}
        \caption{}
        \label{fig:fig1a}
    \end{subfigure}
    \hspace{0.0001\linewidth}
    \begin{subfigure}[b]{0.45\linewidth} 
        \centering
        \includegraphics[width=\textwidth]{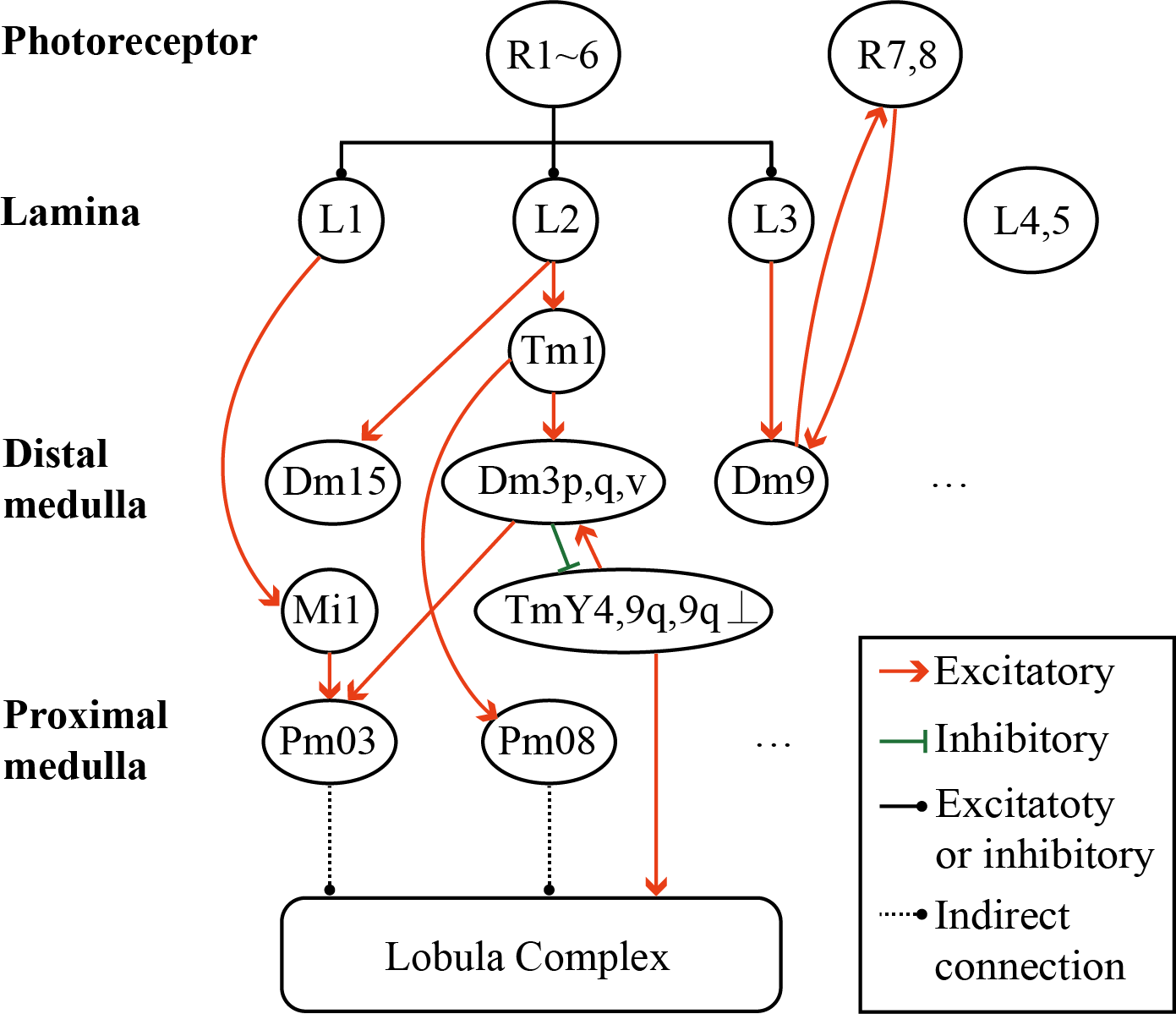}
        \caption{}
        \label{fig:fig1b}
    \end{subfigure}
    \caption{\textbf{Anatomical organisation and connectivity in the fly visual system.} (a) Schematic of the fly optic lobe showing the layered structure of the medulla, including the distal (Dm), serpentine (Sm), and proximal (Pm) sublayers. (b) Simplified connectivity diagram highlighting the visual pathway from photoreceptors (R1-6) through lamina neurons (L1-L3), to the orientation-selective neurons, to the lobula complex. }
\end{figure}

To investigate this, we simulated the visual responses of early-stage neurons (L1-L3) using bar-like stimuli and propagated their activity through a downstream population in the optic lobe modelled with leaky integrate-and-fire (LIF) dynamics, constrained by known synaptic connections \cite{Dorkenwald_Matsliah}. We then analysed the resulting neural population activity, revealing spatially organised orientation tuning reminiscent of cortical orientation maps. Figure \ref{fig:fig1a} provides an overview of the anatomical structure of the \textit{Drosophila} optic lobe, while Figure \ref{fig:fig1b} illustrates the visual connectivity pathway examined in this study.

To summarise, our main contributions are as follows:
\begin{itemize}
    \item We computationally demonstrated, for the first time, spatially coherent orientation maps in the medulla region of an invertebrate visual system.
    \item We identified topological singularities and inter-layer columnar alignment in orientation preference within the distal medulla (Dm) and proximal medulla (Pm) regions.
\end{itemize}
Together, these findings suggest that canonical orientation maps can arise from shared computational principles across species, even in the absence of cortical lamination.

\section{Background and related works}

\paragraph{Orientation maps in mammals.}Orientation selectivity has been extensively studied in the visual cortices of many mammals. At the anatomical level, evidence from mammals indicates that orientation tuning can emerge in neural substrates with vast differences  \cite{Jung_Almasi, Blasdel_Salama, Nauhaus_Benucci}. In rodents, maps often display salt-and-pepper rather than columnar architecture, as observed in the visual cortex, while the retina displays a continuous topographic map of orientation tuning \cite{Vita_Orsi}. In contrast, cats and wallabies display a structured pinwheel-like feature that has been documented in the primary visual cortex \cite{Jung_Almasi_Sun, Jung_Almasi, Merkuleyeva_Lyakhovetskii}, despite differences in cortical evolutionary divergence across mammalian species \cite{Ohki_Chung}. These findings suggest that orientation maps can arise across diverse mammalian species with varying cortical architectures, raising the possibility that such maps are not an exclusive feature of the neocortex but reflect convergent computational motifs. 

\paragraph{Orientation tuning in non-mammals.}Studies across non-mammalian species suggest that structured visual coding arises under minimal anatomical constraints. In pigeons, sharply tuned orientation-selective neurons emerge through feedforward circuits alone, without a laminated cortex \cite{Li_Xiao}. In turtles, despite the absence of fine retinotopy, population activity in the dorsal cortex accurately encodes spatial information \cite{Fournier_Christian}, indicating that a global analysis of the visual scene can arise from distributed representations. Similarly, in zebrafish, orientation-selective neurons have been observed in the optic tectum, and some studies report their laminar segregation and retinotopic organisation \cite{Nevin_Robles, Romano_Pietri}. However, while all three species exhibit orientation-selective neurons, neither has demonstrated the presence of continuous, spatially organised orientation maps comparable to the ones found in the mammalian cortex.

\paragraph{Orientation coding in Drosophila.}In \textit{Drosophila melanogaster}, orientation tuning has been hypothesised based on anatomical structure alone. In particular, \citet{Seung} predicted that functionally distinct neuron types such as Dm3 and TmY may exhibit orientation selectivity, proposing that such responses could arise purely based on wiring alone. To date, the only direct physiological evidence of orientation selectivity comes from earlier studies on T4 and T5 neurons, which exhibit tuning to oriented edges in addition to their well-known direction selectivity \cite{Fisher_Silies}. Beyond these, no other neuron types have been experimentally or computationally confirmed to show orientation selectivity. Moreover, no prior study has computationally demonstrated the emergence of orientation maps in this non-cortical system. Our work fills in this gap by building on these findings to show that coherent, spatially organised orientation maps can emerge in such a system, thus suggesting that species of vastly different evolutionary origin may share common circuit-level principles for encoding visual features.

\section{Methods}
\label{sec:methods}
\begin{figure}
    \centering
    \begin{subfigure}[b]{0.22\linewidth} 
        \centering
        \includegraphics[width=\textwidth]{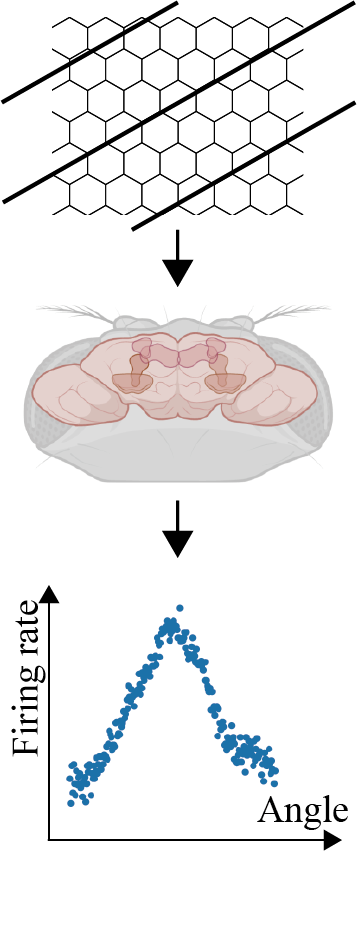}
        \caption{}
        \label{fig:fig2a}
    \end{subfigure}
    \begin{subfigure}[b]{0.76\linewidth}
        \centering
        \begin{subfigure}{\linewidth}
            \centering
            \includegraphics[width=\textwidth]{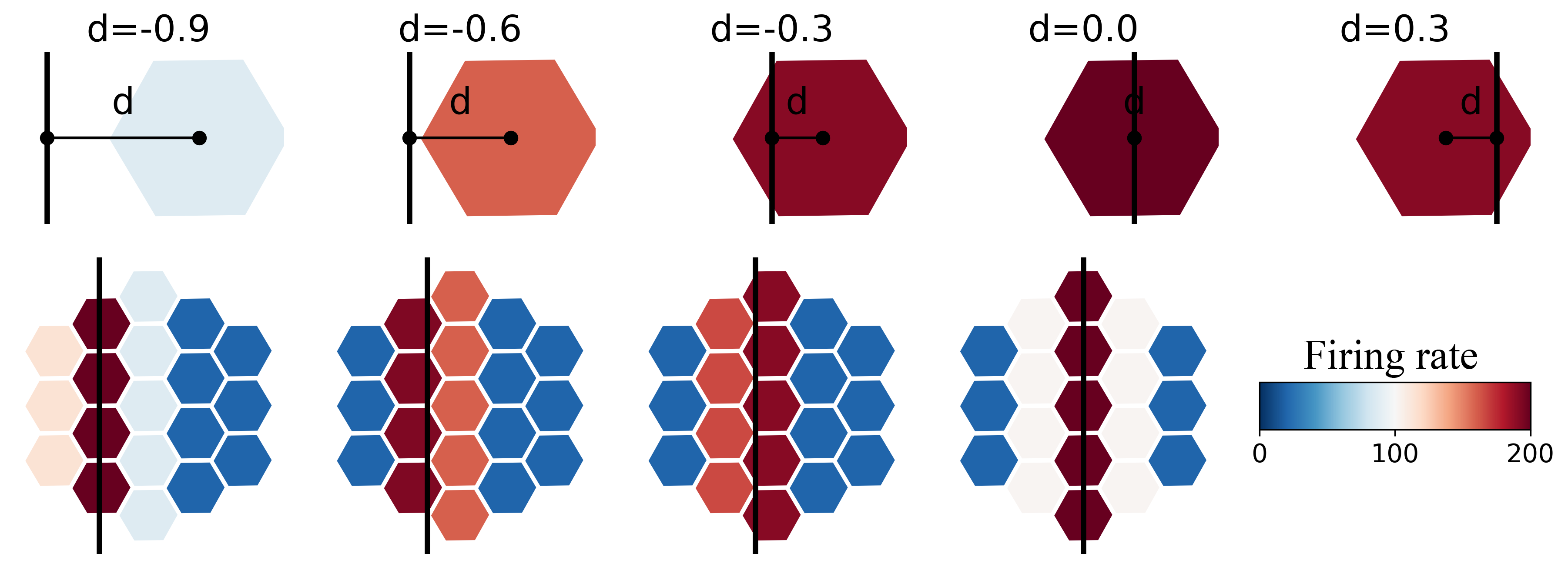}
            \caption{}
            \label{fig:fig2b}
        \end{subfigure} \\
        \begin{subfigure}{0.69\linewidth}
            \centering
            \includegraphics[width=\textwidth]{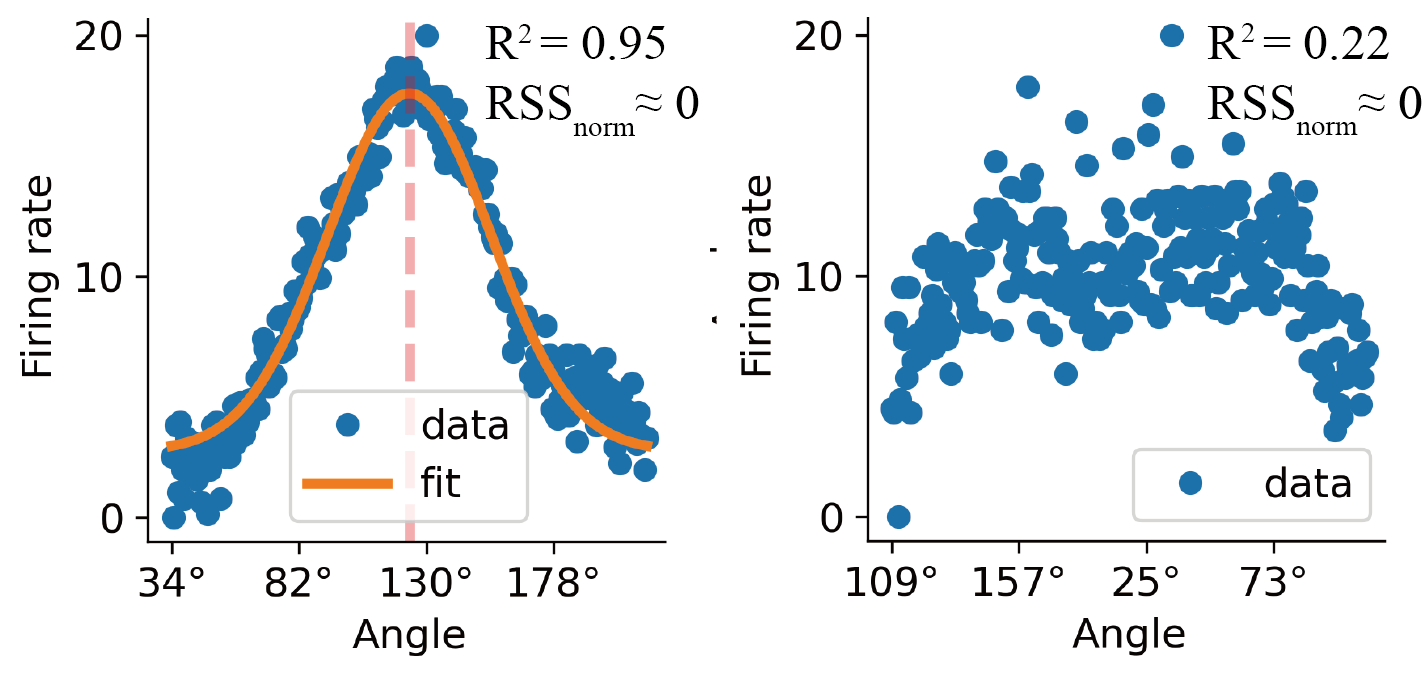}
            \caption{}
            \label{fig:fig2c}
        \end{subfigure}
        \hspace{0.001em}
        \begin{subfigure}{0.28\linewidth}
            \centering
                \includegraphics[width=\textwidth]{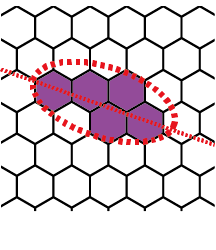}
                \caption{}
            \label{fig:fig2d}
        \end{subfigure}
    \end{subfigure}
    
    \caption{\textbf{Overview of the simulation framework and analysis methods.} (a) Schematic of the simulation pipeline. Poisson spike trains were used to simulate grating stimuli input to L1-L3 neurons. Each neuron's orientation tuning is quantified by its firing rate across stimulus angles. (b) Structured stimulus input to L1-L3 neurons across different OFF-bar positions (black line). Top: Each panel shows a hexagon representing an ommatidium. Bottom: 2D heatmap of firing activity across the simplified hexagonal lattice of ommatidia. Each hexagon represents one ommatidium, colour-coded by its corresponding L1-L3 firing rate. $d$ represents the distance to the centre of the ommatidia. Please note that, even when the bar stimulus lies outside the geometric boundary of an ommatidium, a measurable response is still observed due to the broader receptive field (~5\degree - 8\degree) relative to the ommatidial spacing (4.6\degree), resulting in spatial overlap of sensitivity across neighbouring units\cite{BORST2009R36}. (c) Gaussian turning curves of two neurons: the left panel indicates a well-fit neuron (orientation-selective), and the right panel indicates a neuron classified as not orientation-selective. (d) Structural prediction of preferred orientation angles based on upstream columnar connectivity. The red dashed lines indicate the best-fit ellipse drawn through these spatial locations.}
    \label{fig:enter-label}
\end{figure}

\paragraph{LIF model.} We implemented a leaky integrate-and-fire (LIF) framework to simulate the spiking dynamics of neurons in the \textit{Drosophila} visual system (see Appendix \ref{sec:appendix-lif-design}) \cite{Shiu_Sterne}. Using the full adult fly brain (FAFB) connectome, we constructed a network in which neurons are labelled and connected according to their anatomical synapses \cite{Buhmann_Sheridan, Dorkenwald_Matsliah, Heinrich_Funke-key, Zheng_Lauritzen}. The model encompasses the complete adult \textit{Drosophila} connectome, including 138,639 neurons and 1,508,983 synapses. The FAFB connectome provides a complete and cell-resolved reconstruction of the \textit{Drosophila} brain \cite{Zheng_Lauritzen}, and has become a foundational resource for structural annotation \cite{Schlegel_Yin}, functional inference \cite{Walker_2025, Cowley_2022}, and whole-brain spiking simulations validated against behaviour \cite{Shiu_Sterne}. In our model, visual stimuli were simulated as Poisson spike trains injected into lamina inputs (L1-L3) in the right eye of the \textit{Drosophila}, and the resulting activity was propagated through the whole brain connectome using the embedded LIF framework, allowing us to monitor orientation tuning in downstream visual neurons (Figure \ref{fig:fig2a}).

\paragraph{Stimuli.} 
We focused on the lamina neurons L1-L3, which are visual columnar neurons - retinotopically organised neurons that are associated with an individual ommatidium in the compound eye, such that each visual column contains a dedicated copy of the neuron \cite{Nern_2025}. Unlike other columnar neurons, L1-L3 receive direct input from photoreceptors R1-R6, which capture brightness change across the visual field. The ommatidia themselves are arranged in a hexagonal lattice, forming a precise sampling grid over visual space. For this study, we applied input directly to L1-L3 rather than R1-R6, as the connectomic dataset provides visual columnar mappings for L1-L3 but not for photoreceptors \cite{Buhmann_Sheridan, Dorkenwald_Matsliah, Eckstein_Bates, Heinrich_Funke-key, Lin_Yang, codex, Mastsliah_Yu, Schlegel_Yin, Zheng_Lauritzen, Nern_2025}. Naively, one can let an L1-L3 neuron receive a fixed strength of Poisson input regardless of the location of the OFF-bar on the receptive field (RF) of the corresponding ommatidium. However, this ignored spatial gradients across the RF. To better reflect spatial structure, the input stimuli were modelled by computing the distance $d$ from each ommatidium to the bar and making the input strength depend on $d$ based on the calcium imaging data \cite{Ketkar_Burak} (Appendix ~\ref{sec:appendix-stimulus}). Figure ~\ref{fig:fig2b} illustrates how the distance of the OFF-bar modulates input strength. In the top row, each panel shows a single ommatidium, colour-coded by its corresponding L1-L3 firing rate as a function of $d$ from the OFF-bar (black line). Firing rates increased as the bar approached the centre of the ommatidia, peaking when aligned and tapering off with distance. The bottom row provides a simplified illustration of the 2D activation pattern across the hexagonal lattice for several OFF-bar positions.  

\paragraph{Preferred orientation.}To quantify the orientation selectivity of the monitored neurons, we fit a Gaussian function to their firing responses across different stimulus orientations. Unlike a standard Gaussian, which assumes a linear domain, our model accounts for the \(180\degree\) periodicity of angle space by computing the minimal circular distance between the stimulus orientation and the neuron's preferred orientation (Figure \ref{fig:fig2c}; see Appendix~\ref{sec:circular_gaussian_function}). A fit was considered "good" if it satisfied two criteria: a coefficient of determination \(R^2 \geq 0.7\) and a normalised residual sum of squares \(RSS_{norm} \leq 0.4\) -- indicating both high explanatory power and low relative error (Figure \ref{fig:fig2c}). Representative examples of Gaussian fits that are considered good, poor and near threshold are illustrated in Figure \ref{fig:gaussian_1a}. The distribution of $R^2$ across neurons and the relationship between $R^2$ and $RSS_{norm}$ are illustrated in Figure \ref{fig:gaussian_1b}, \ref{fig:gaussian_1c}. Unless otherwise noted, the preferred orientation of a neuron in this paper refers to its orientation preference as determined by this Gaussian fit.

\paragraph{Structural prediction.}In this paper, structured prediction refers to the predicted preferred orientation angle of a neuron by its direct upstream neurons, similar to the receptive field approach used by \citet{Seung}. Specifically, we identify each neuron's upstream partners that are also columnar neurons using known synaptic connectivity. We mapped these upstream neurons to their corresponding columns in the compound eye, and a best-fit ellipse is drawn around their spatial locations (Figure \ref{fig:fig2d}). The orientation of the ellipse's major axis, measured relative to the y-axis, is taken as the predicted angle for the downstream neuron.

\section{Results}
\subsection{Orientation selectivity}
\begin{figure}[tp]
    \centering
    \begin{subfigure}[b]{\linewidth} 
        \centering 
        \includegraphics[width=\textwidth]{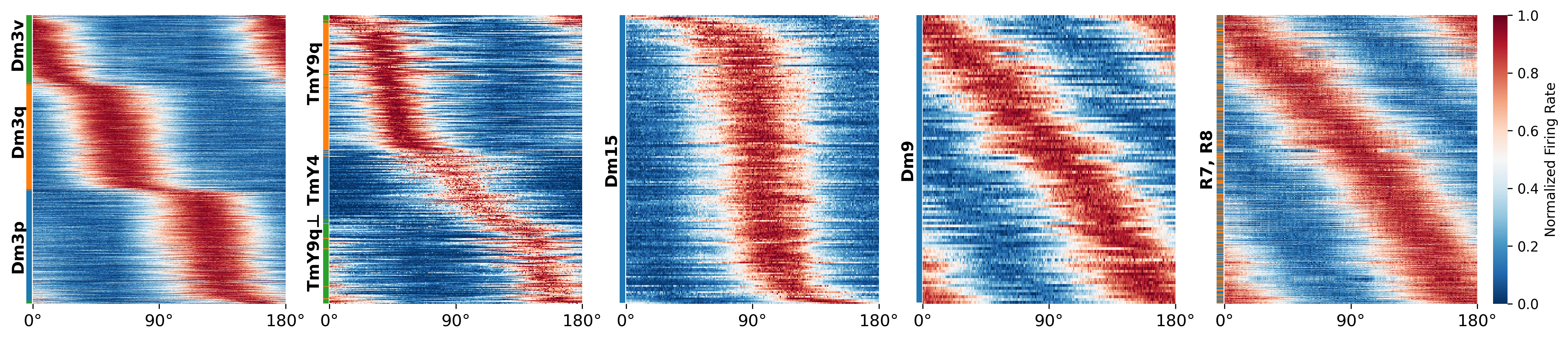}
        \caption{}
        \label{fig:fig3a}
    \end{subfigure} \\
    \begin{subfigure}[b]{0.74\linewidth}
        \centering 
        \includegraphics[width=\textwidth]{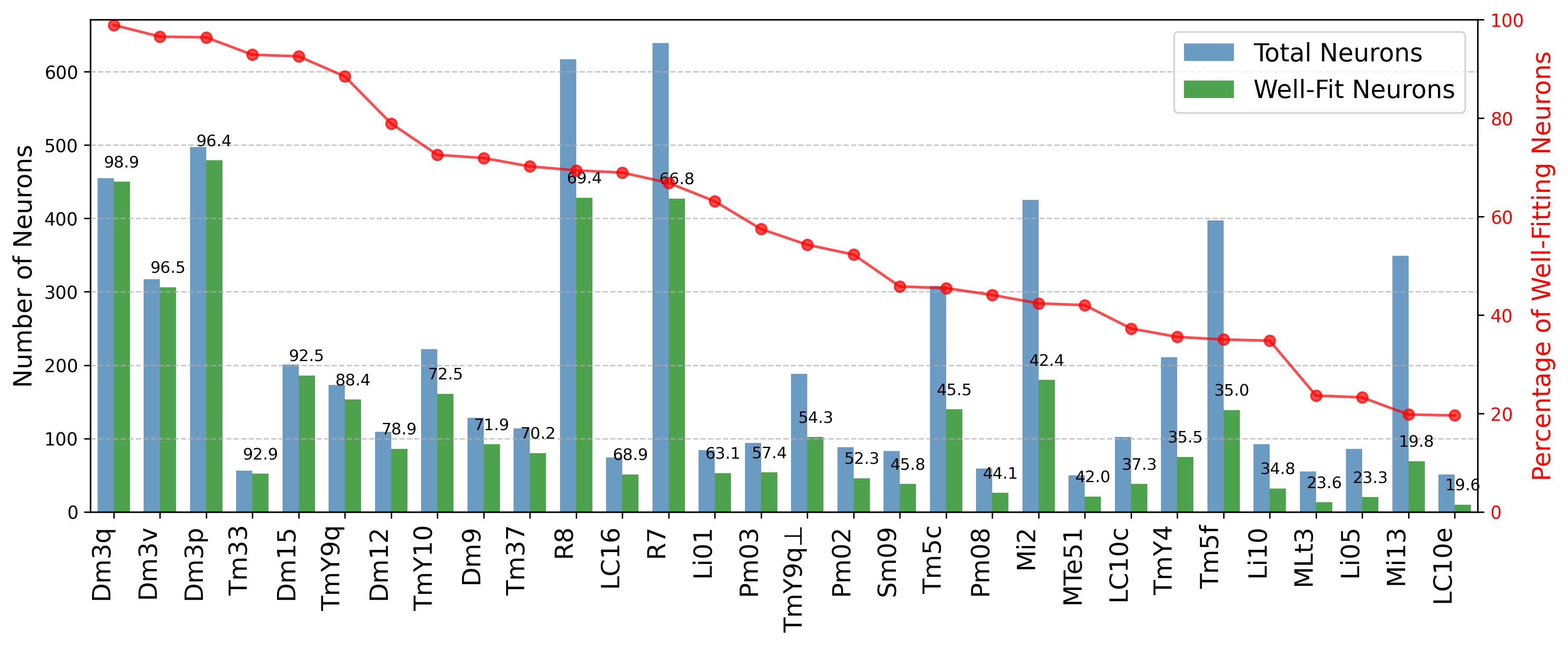}
        \caption{}
        \label{fig:fig3b}
    \end{subfigure}
    \begin{subfigure}[b]{0.25\linewidth}
        \centering 
        \includegraphics[width=\textwidth]{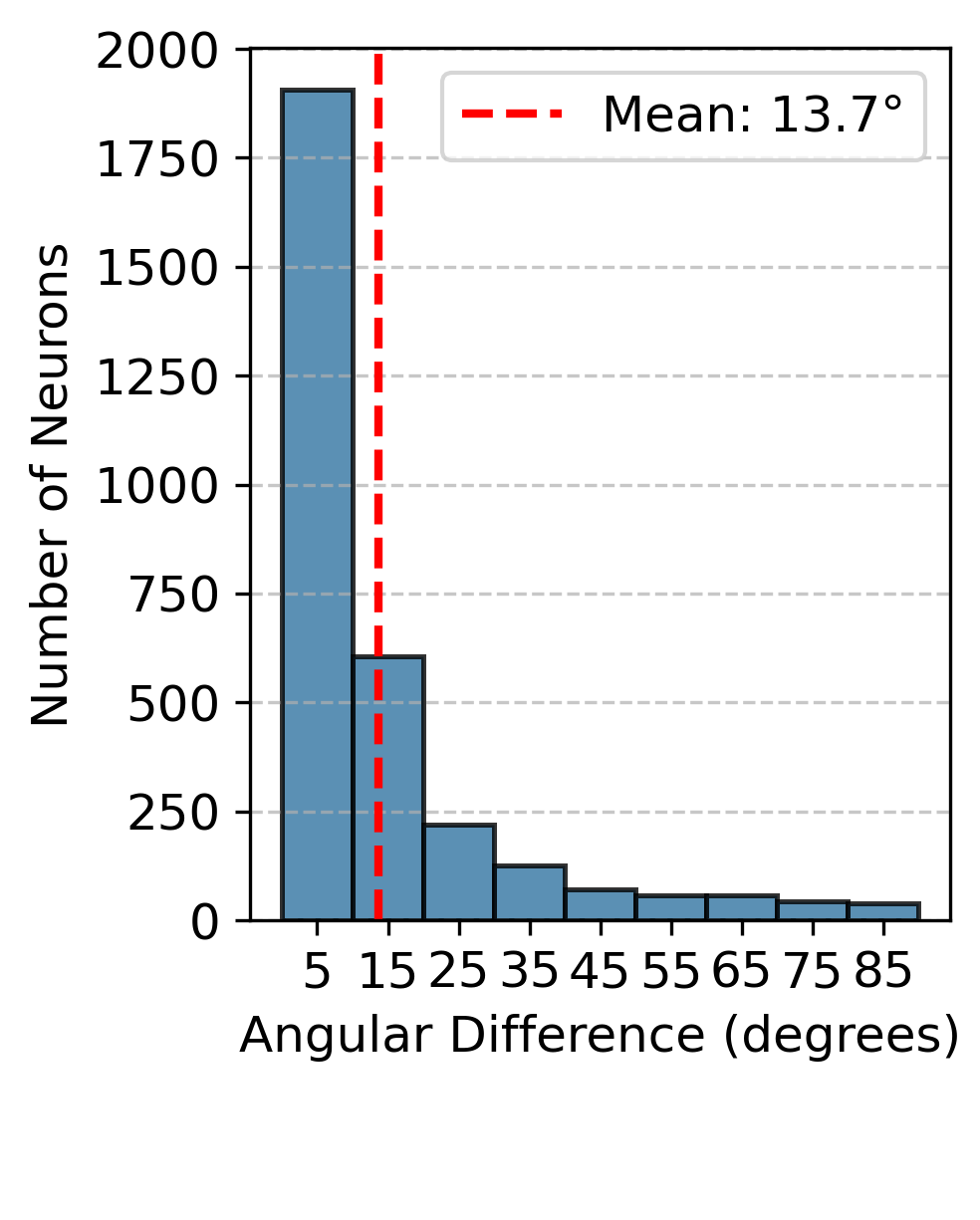}
        \caption{}
        \label{fig:fig3c}
    \end{subfigure}
    \caption{\textbf{Orientation tuning and structural prediction accuracy.} (a) Heatmaps showing the normalised firing rates of various neurons across different angular orientations. (b) Proportion of neurons with well-fit orientation tuning across types. Green bars indicate the subset well-fitted by a circular Gaussian model (see methods). Blue bars indicate the total number of neurons in the right optic lobe of the respective type. (c) Distribution of absolute differences between each neuron's preferred orientations predicted by LIF simulation and by dendritic structure.}
\end{figure}

We first evaluated whether neurons in the \textit{Drosophila} optic lobe exhibit robust orientation selectivity. Using our LIF-based simulation framework, we presented bar-like stimuli across orientations and recorded their firing rates. We then fitted each neuron's response profile using a circular Gaussian model (see Methods). For neurons that were well-fit, we normalised firing rates to their peak responses and visualised the resulting heatmap of preferred orientations. Neurons were initially sorted via hierarchical clustering, revealing structured patterns in tuning preferences. Upon observing a diagonal-like structure - suggestive of a continuous orientation gradient - we further sorted neurons by the ascending position of their peak response, which highlighted a clearer gradient in orientation tuning (Figure \ref{fig:fig3a}). Our findings demonstrated that \textit{Drosophila} neurons exhibit robust orientation selectivity, with evidence suggesting that orientation maps in this species may exist. The percentage of well-fit neurons relative to the total neurons in the type for a subset of neurons (Dm3v, Dm3p, Dm3q, TmY4, TmY9q, TmY9q\(^\perp\)) aligns well with Seung's prediction \cite{Seung}, indicating that orientation selectivity arises directly from the spatial arrangement of dendritic inputs (left two panels of Figure \ref{fig:fig3a}).

Moreover, we found that several other neuron types (e.g., Dm15, Tm33, TmY10) also exhibited tuning properties (Figure \ref{fig:fig3a}). To quantify this selectivity, we classified a neuron type as orientation selective if over $40\%$ of its neurons were well fit (Figure \ref{fig:fig3b}; see Methods). These results revealed that most orientation-selective neurons were located within the medulla layer, highlighting it as a key site for early orientation processing in the \textit{Drosophila} visual system. To further investigate the tuning mechanism and verify our hypothesis that orientation tuning could be directly influenced by its dendritic inputs, we quantified the preferred orientation of each neuron by computing the structural prediction of the neuron (Figure \ref{fig:fig2a}; see Methods). We then compared it to the neuron's preferred orientation using absolute difference. Across the population analysed, the mean absolute difference sits at \(13.7\degree\) (Figure \ref{fig:fig3c}), indicating a robust correspondence between dendritic geometry and orientation preference. Notably, T4 and T5 neurons, despite being known for their direction and orientation selectivity, did not meet our selection criteria in this analysis. Our results showed that most T4 and T5 neurons had firing rates close to zero. A few neurons that had higher firing rates exhibited orientation selectivity (Figure \ref{fig:ex_fig2}). This was likely due to the stimulus design: the static input may have been insufficient to strongly activate these neurons. 

Intriguingly, while analysing the orientation tuning patterns across neuron populations, we incidentally observed that photoreceptor R7 and R8, classically known for detecting colours \cite{Mastsliah_Yu}, also exhibit orientation selectivity (Figure \ref{fig:fig3b}). This unexpected finding raises the question of how such tuning arises in primary sensory neurons. We investigated the connectivity of the neurons and found that Dm9, R7 and R8 formed a closed recurrent loop that results in R7 and R8 adopting the same orientation preferences as Dm9 (Figure \ref{fig:fig1b} and right two panels of Figure \ref{fig:fig3a}), which itself receives input from L3. This closed recurrent loop between Dm9, R7 and R8, leading to synchronised orientation preferences, could provide new insights into how feedback within small circuits may amplify or stabilise neural responses. While this was not the primary hypothesis, it raises interesting questions about the role of recurrent circuits in orientation selectivity. Further exploration of this feedback mechanism in the future could contribute to our understanding of how local orientation selectivity is maintained or enhanced within neural networks.

\subsection{Orientation maps}
\begin{figure}[t]
    \centering
    \begin{subfigure}[b]{0.31\linewidth} 
        \centering
        \includegraphics[width=\textwidth]{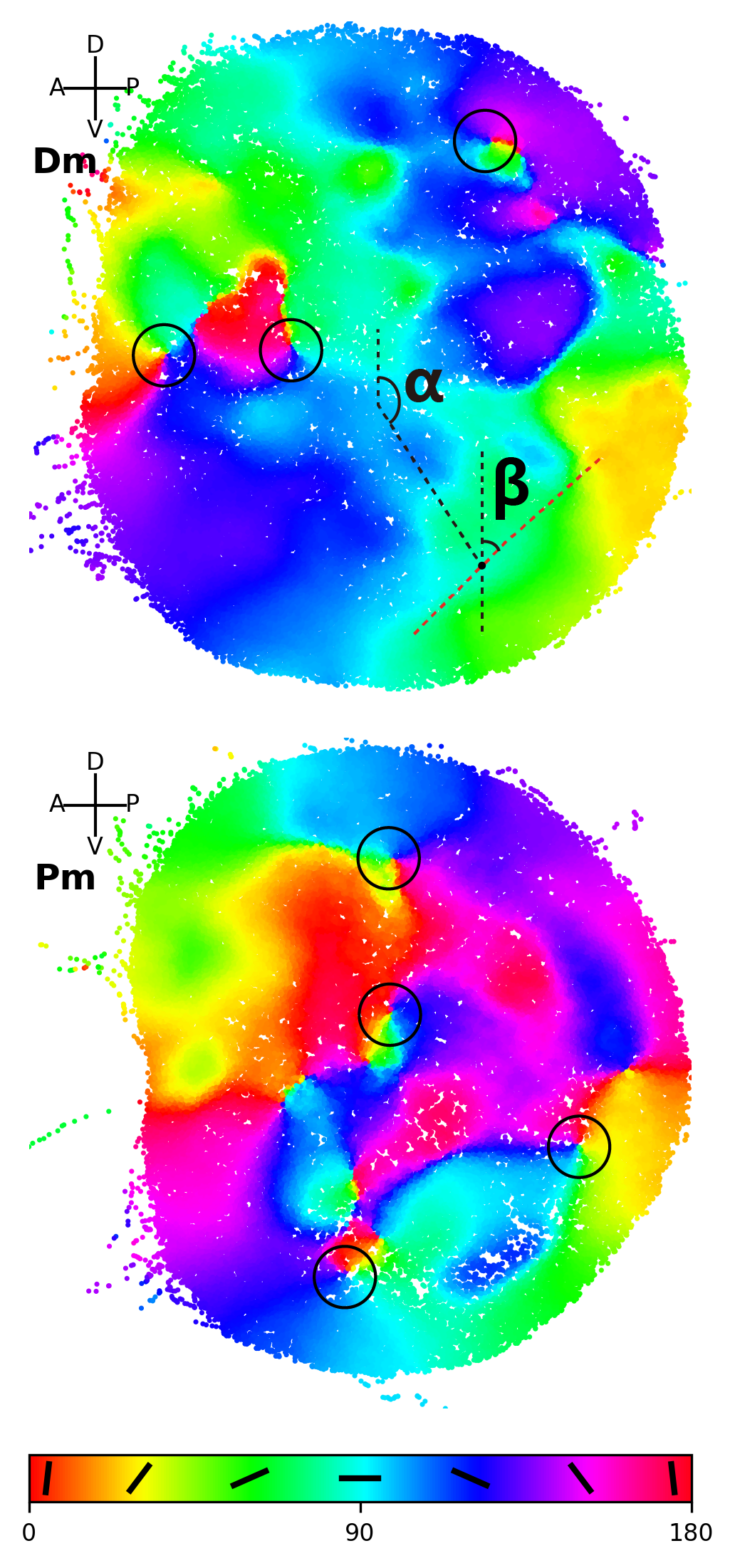}
        \caption{}
        \label{fig:fig4a}
    \end{subfigure}
    \begin{subfigure}[b]{0.31\linewidth} 
        \centering
        \includegraphics[width=\textwidth]{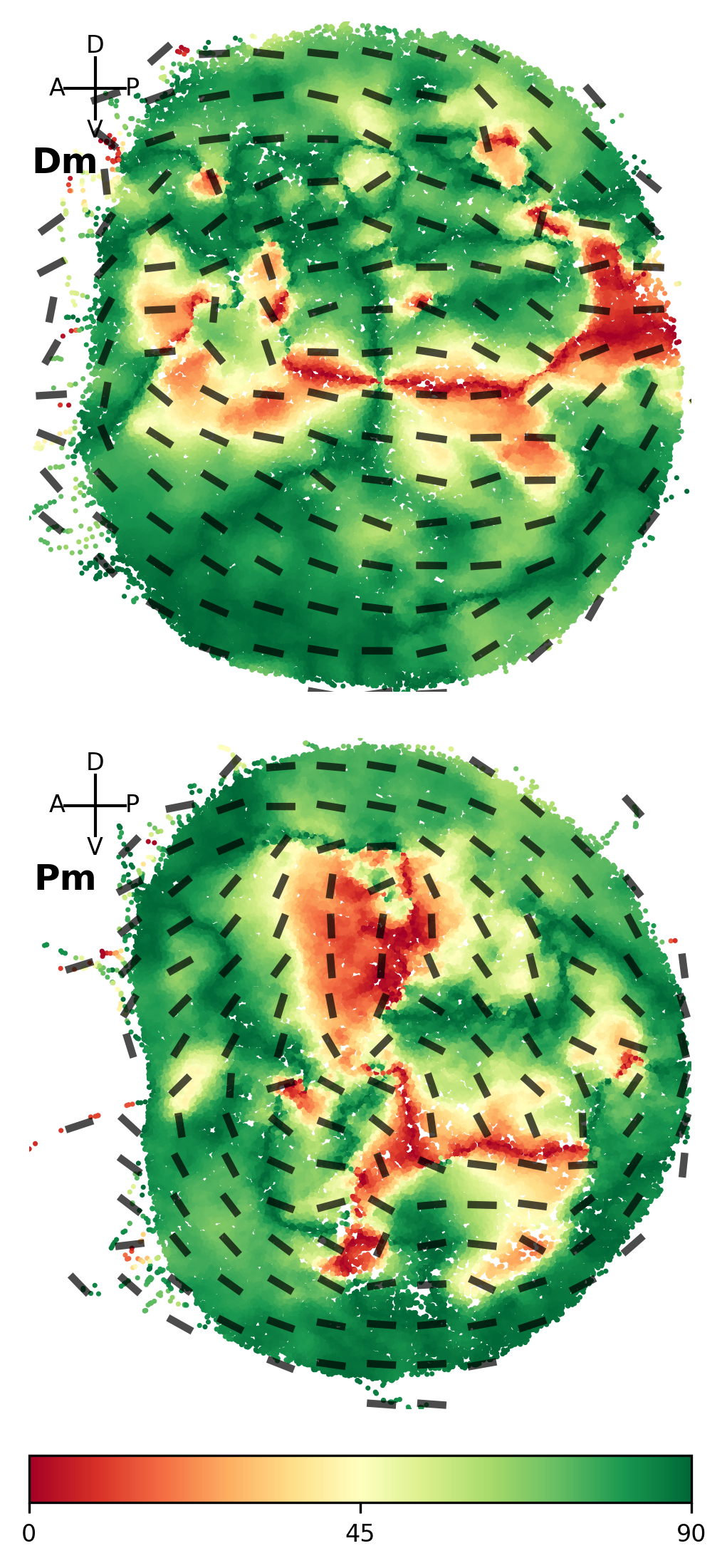}
        \caption{}
        \label{fig:fig4b}
    \end{subfigure}
    \begin{subfigure}[b]{0.25\linewidth} 
        \centering
        \begin{subfigure}{\linewidth}
            \includegraphics[width=\textwidth]{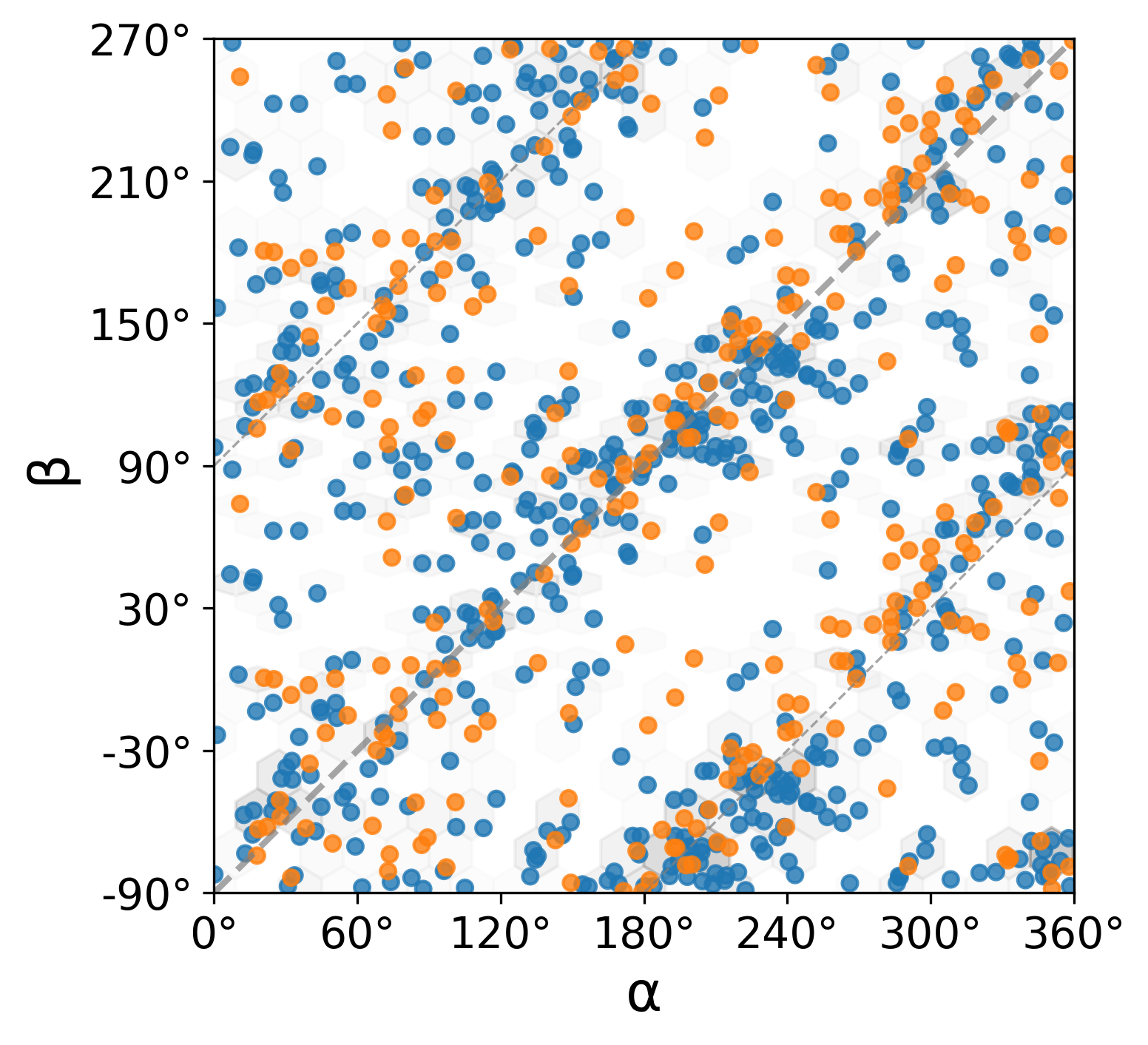}
            \caption{}
            \label{fig:fig4c}
        \end{subfigure}
        \begin{subfigure}{\linewidth}
            \includegraphics[width=\textwidth]{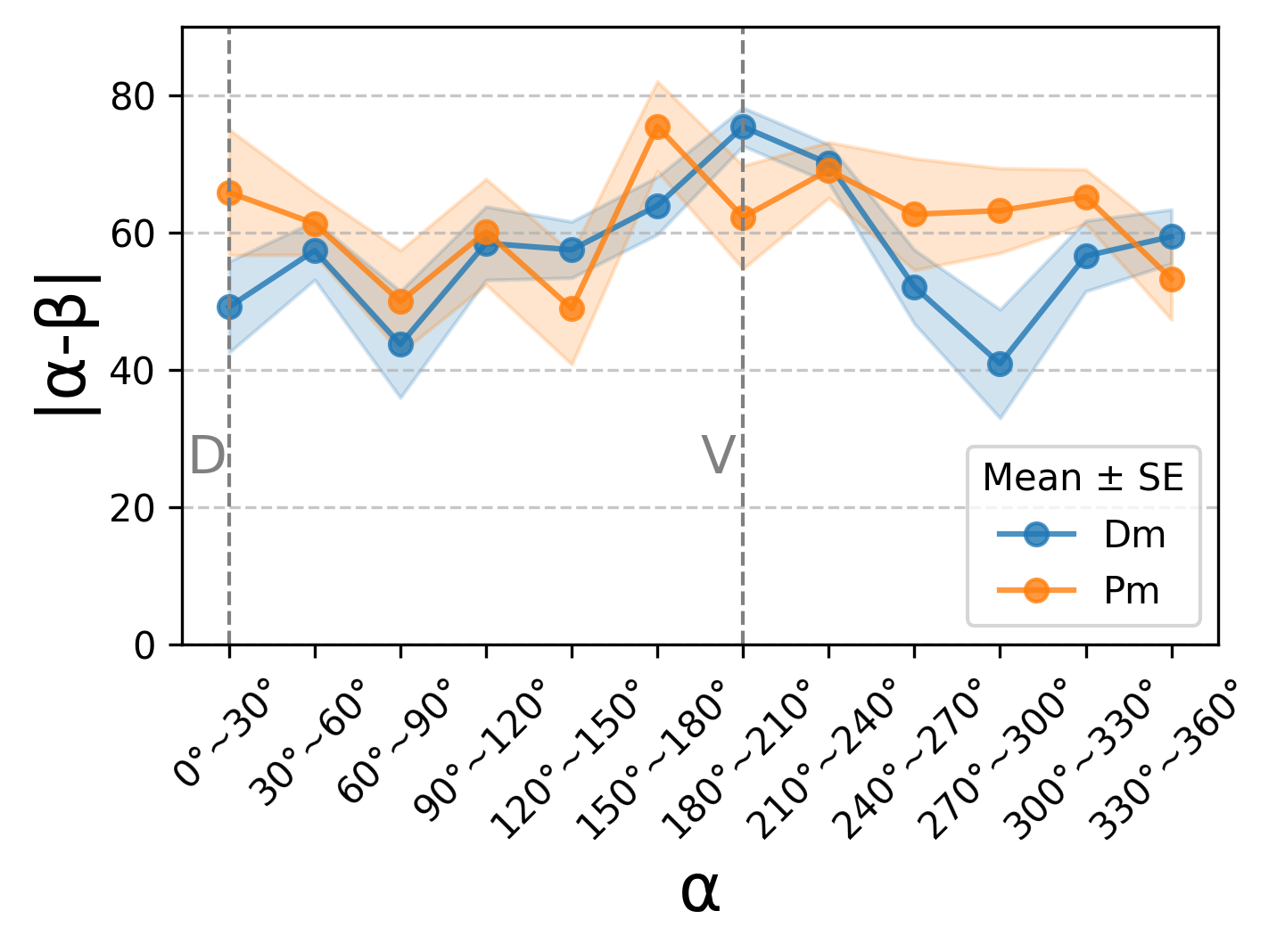}
            \caption{}
            \label{fig:fig4d}
        \end{subfigure}
        \begin{subfigure}{\linewidth}
            \includegraphics[width=\textwidth]{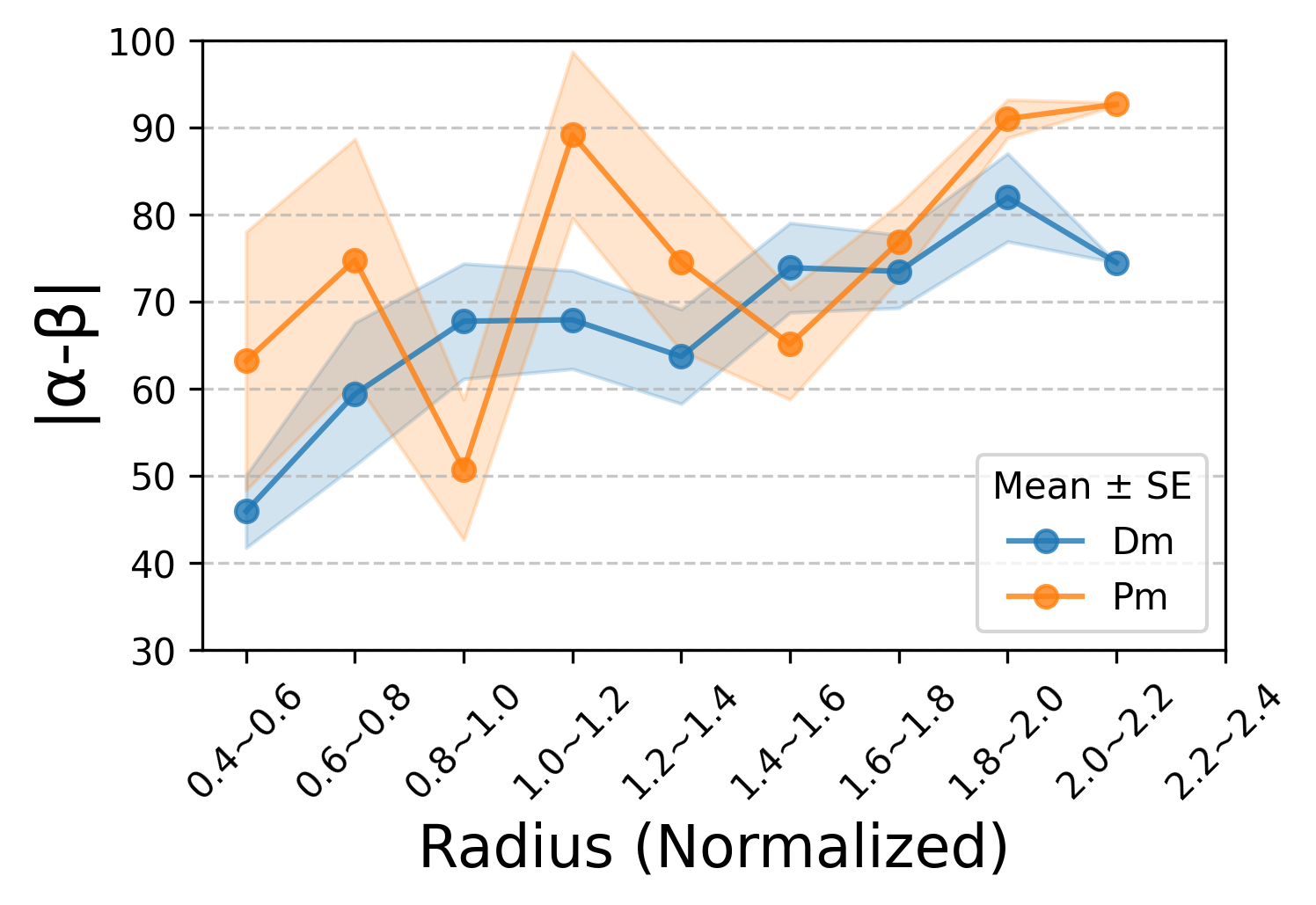}
            \caption{}
            \label{fig:fig4e}
        \end{subfigure}
    \end{subfigure}
    \caption{\textbf{Spatial and orientation map in the optic lobe.} (a) The smoothed heatmap of the optic lobe coloured by its preferred orientations at each layer of the optic lobe. The circles highlight the observed pinwheel structures. $\alpha$ denotes the position angle where dorsal-ventral represents $0\degree$ and $\beta$ denotes the preferred orientation of the neurons (see Methods). (b) Vector field plots of the preferred orientation in the Dm (top) and Pm (bottom) layers. Each line represents the preferred orientation of a neuron, and the background heatmap encodes the absolute angular difference between the neuron's $\alpha$ and $\beta$. Green regions indicate strong tangential alignment ($\Delta \approx 90\degree$), red regions indicate radial alignment ($\Delta \approx 0\degree$). (c) Scatter plot (position angle, preferred orientations) pairs are displayed with coordinate axes shifted for visualisation (y-axis: \(-90\degree\) to \(270\degree\), x-axis: \(0\degree\) to \(360\degree\)). (d) Relationship of $|\alpha - \beta|$ and $\alpha$. D: dorsal area; V: ventral area. (e) Relationship between $|\alpha - \beta|$ and radial distance from the centre.}
\end{figure}
To evaluate whether orientation selectivity is spatially organised in the fly visual system, we focused on the medulla, a key visual processing region that receives direct input from L1-L3 neurons and contains the highest density of neurons with strong orientation selectivity (Figure \ref{fig:fig3b}). Across the medulla sublayers - which are anatomically divided into distal (Dm), serpentine (Sm) and proximal (Pm) layers, shown in Figure \ref{fig:fig1a}, we identified 1710 neurons in the Dm layer, of which 1245 belonged to the Dm3v, Dm3q and Dm3p subtypes. Additionally, we identified 143 well-fit neurons in the Pm layer and 63 in the Sm layer, based on the Gaussian tuning criteria (see Methods). It was found that neurons in all three sublayers exhibited clustered patterns - neurons with similar preferred orientations tended to cluster together (Figure \ref{fig:ex_fig3}). We then focused on the Dm and Pm layers for detailed spatial analysis, as these layers contained a sufficient number of neurons spanning the full spatial extent of the medulla (Figure \ref{fig:ex_fig3}). 

We first visualised the spatial organisation of preferred orientation by flattening the 3D morphology of the neurons in the Dm and Pm layers onto layer-specific 2D coordinate systems. To achieve this, we computed a best-fit plane through the full set of 3D skeleton coordinates from each layer, providing a reference frame for spatial alignment and visualisation. This approach preserved the columnar layout while minimising distortions introduced by the \textit{Drosophila's} optic lobe curvature. To enable cross-layer comparisons, a scaling transformation was applied to normalise for differences in physical size and curvature between the Dm and Pm regions. This projection preserves relative topographic relationships while enabling 2D spatial smoothing and angular comparisons. We colour-coded the neurons into their preferred orientation across Dm and Pm layers using unsmoothed maps to examine the spatial distribution of preferred orientations (Figure \ref{fig:ex_fig3}). These maps revealed a clear spatial clustering of similar orientation preferences, in contrast to a salt-and-pepper pattern in the rodent V1 area. 

To better visualise the global structure, we applied local spatial smoothing by computing the circular mean of preferred orientations within a circular neighbourhood of fixed radius $r$ around each neuron. Specifically, for each neuron in the 2D projected plane, we identified all neurons located within a disk of radius $r \approx 5 \times 10^3~\mathrm{nm}$ and computed the circular mean of their orientation preferences. This smoothing preserved mesoscale spatial patterns while reducing local variability. The resulting heatmap produced orientation maps shown in Figure \ref{fig:fig4a}. These smooth maps reveal pinwheel-like singularities - points around which orientation preference rotates continuously (the circled points in Figure \ref{fig:fig4a}) - indicating a degree of spatial coherence previously unreported in invertebrate visual systems.

Initial observations of Figure \ref{fig:fig4a} revealed that both layers exhibited a roughly centrosymmetric distribution of preferred orientations. To characterise this structure more precisely, we defined a polar coordinate system for each layer, with the origin set at the centroid of the neuron population, computed by the mean of the spatial coordinates, and using the V-D (bottom-top) direction as the axis. This allowed us to compute a spatial position angle $\alpha$ for each neuron and directly compare it to its preferred orientation $\beta$, derived from the smoothed map in Figure \ref{fig:fig4a}. Building on this smoothed representation, we treated each neuron's orientation as representative of its surrounding neighbourhood, based on the same local averaging kernel described earlier. To quantify alignment, we computed the angular difference $\Delta = |\alpha - \beta|$ for each neuron and projected it onto a 2D heatmap (Figure \ref{fig:fig4b}), where $\Delta = 90\degree$ indicates tangential alignment and $\Delta = 0\degree$ indicates radial alignment. This visualisation revealed a clear trend: neurons in peripheral regions tended to prefer tangential orientations ($\beta \approx \alpha + 90\degree$). These findings suggest that orientation tuning in the medulla is not randomly distributed, but spatially organised relative to each neuron's anatomical position.


We then further examined the relationship between $\alpha$ and $\beta$ for all neurons in Dm and Pm layers. Here, each neuron's position angle was computed as the circular mean of the angular positions of all its voxels in the 2D projection. The resulting scatter plot showed a clear diagonal band, consistent with a systematic offset where $\beta \approx \alpha + 90\degree$, supporting the presence of tangential alignment (Figure \ref{fig:fig4c}). To quantify this relationship, we computed $\Delta$ and examined how it varied across spatial dimensions. Across angular bins, the $\Delta$ remains relatively stable with minor regional differences (Figure \ref{fig:fig4d}). In contrast, alignment improved monotonically with radial distance from the centre (Figure \ref{fig:fig4e}), indicating stronger alignment at the periphery. Although perfect tangential alignment would yield $\Delta \approx 90\degree$, the observed means approaching \(90\degree\) suggest a systematic but approximate offset. We speculate that radial and tangential recognition neurons organise an overall pattern in the medulla corresponding to the horizontal direction. This pattern is especially evident in the Dm layer, suggesting a non-uniform organisation of orientation preferences, which may play a key role in the previously reported recognition of horizons and horizontal objects by the biological flight control system \cite{straw_visual_2010}.


This finding suggests that orientation tuning in \textit{Drosophila's} optic lobe is not randomly distributed but exhibits a structured, position-dependent organisation. To access the contribution of upstream visual pathways to orientation selectivity, we performed a targeted ablation analysis (Appendix \ref{sec:ablation}). Silencing Mi neurons abolished tuning across both Dm and Pm layers, while silencing Tm neurons produced selective loss in Pm but minimal effect in Dm, suggesting that Mi neurons provide essential input for orientation tuning throughout the optic lobe. Consistent with this observation, an analysis of synaptic connectivity by orientation preference (Appendix \ref{sec:orientation_connectivity}) revealed that excitatory neurons with similar tuning are more likely to form recurrent connections, supporting topographic organisation and collinear facilitation as proposed by \citet{Seung}. In contrast, inhibitory connections displayed distinct off-diagonal structure, suggesting selective cross-orientation motifs that may sharpen tuning and maintain balance within the network. Peripheral neurons preferentially align their orientation selectivity tangentially relative to the centre of the optic lobe, reminiscent of contour-aligned representations seen in higher animals.

\subsection{Orientation columns}
\begin{figure}[tp]
    \centering
    \begin{subfigure}[b]{0.76\linewidth}
        \centering
        \begin{subfigure}{\linewidth}
            \centering
            \includegraphics[width=\linewidth]{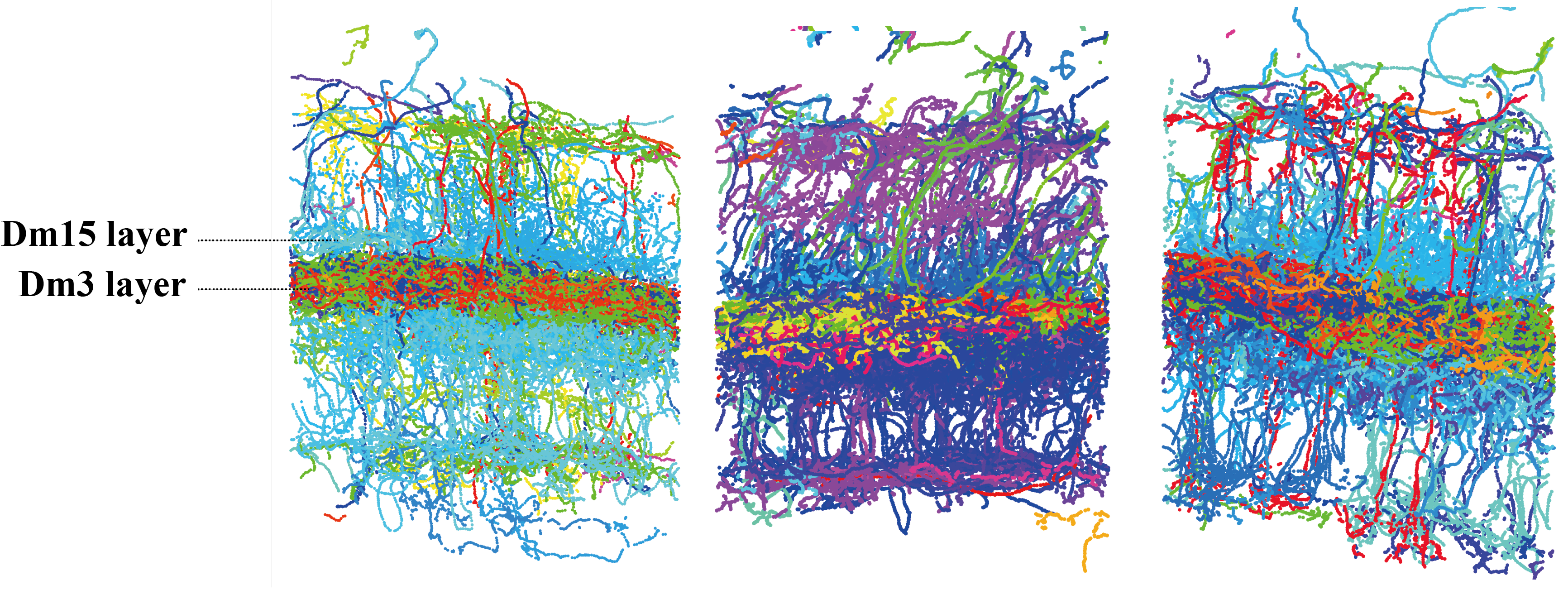}
            \caption{}
            \label{fig:fig5a}
        \end{subfigure} \\
        \begin{subfigure}{\linewidth}
            \includegraphics[width=\linewidth]{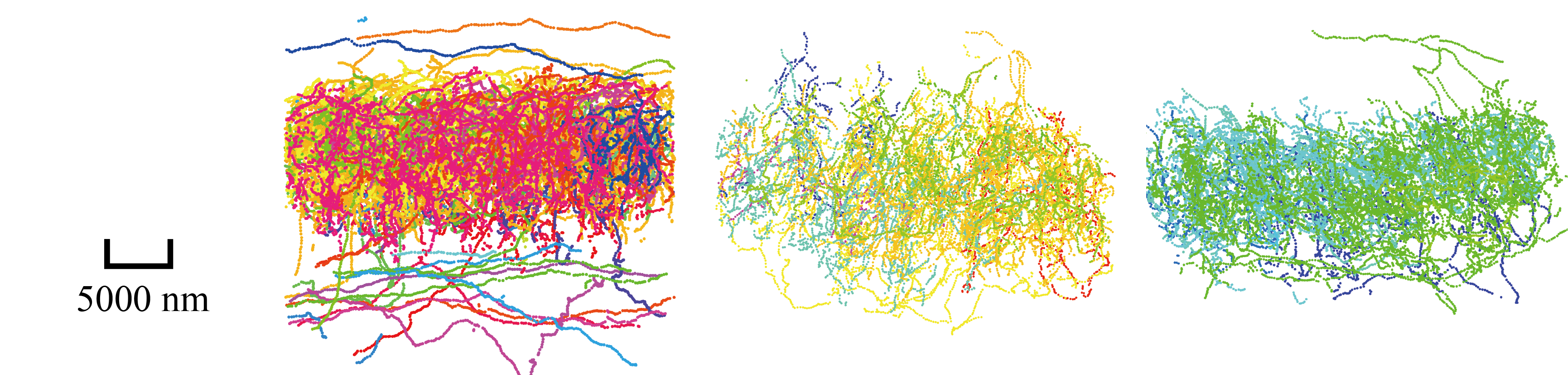}
            \caption{}
            \label{fig:fig5b}
        \end{subfigure}
    \end{subfigure}
    \begin{subfigure}[b]{0.23\linewidth}
        \begin{subfigure}{\linewidth}
            \includegraphics[width=\linewidth]{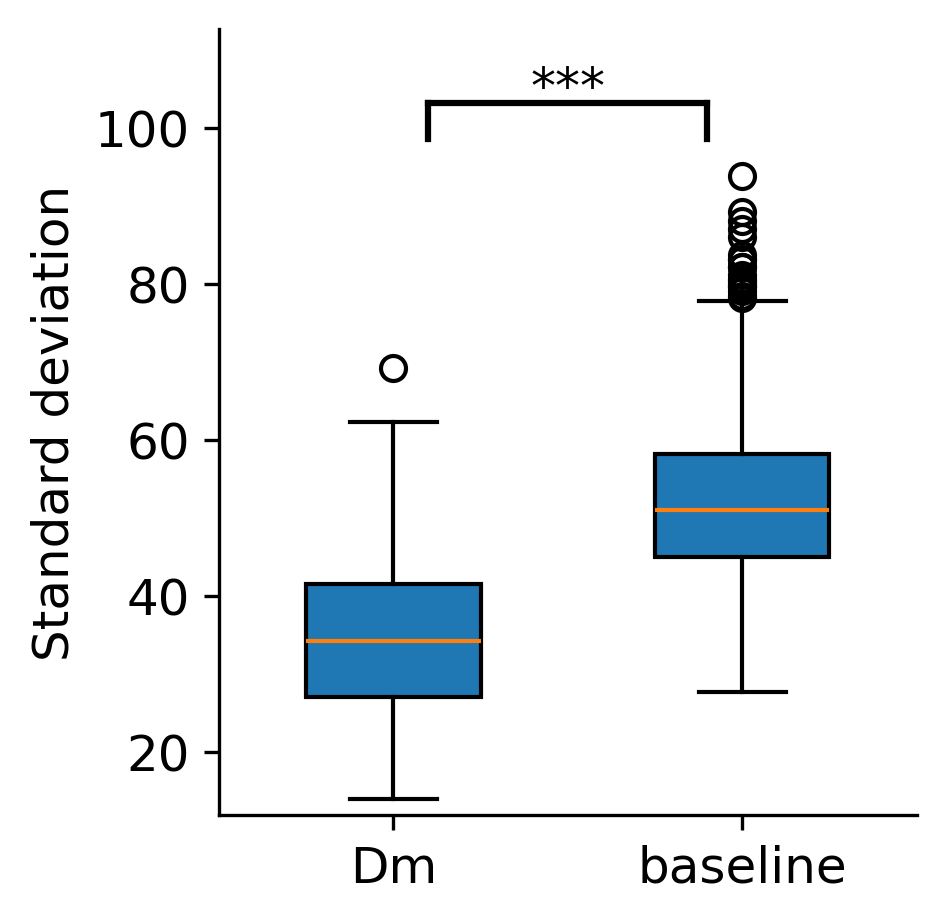}
            \label{fig:fig5c}
        \end{subfigure} \\
        \begin{subfigure}{\linewidth}
            \includegraphics[width=\linewidth]{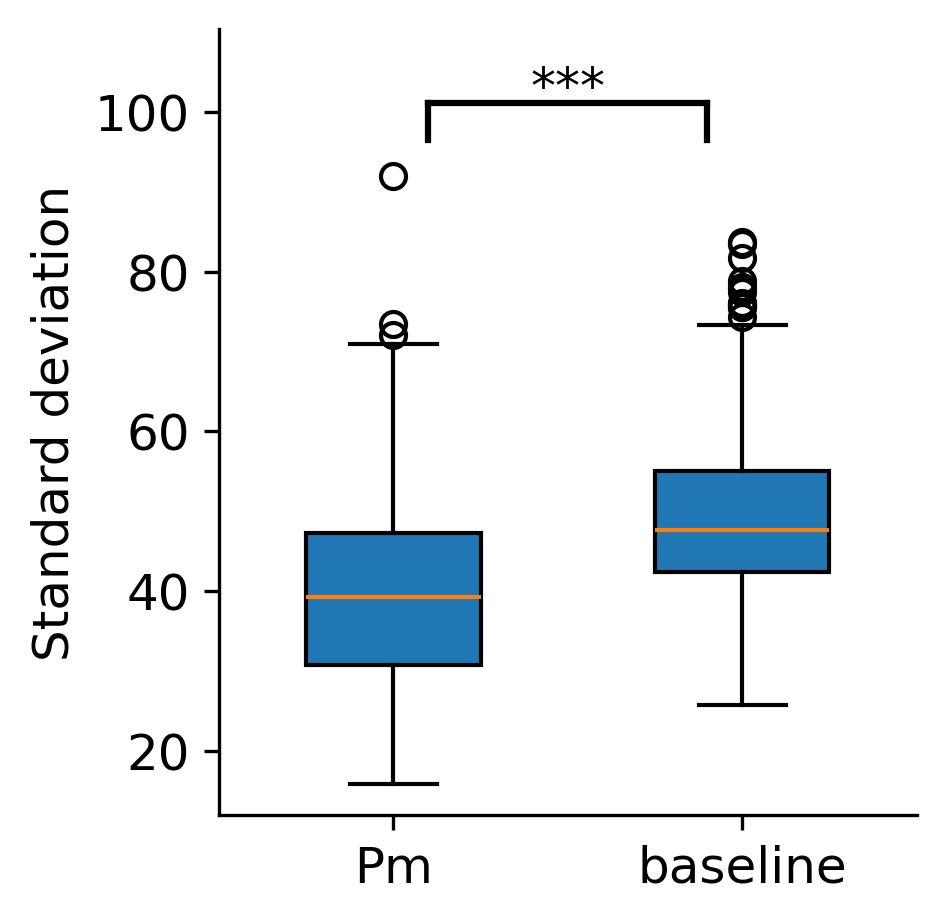}
            \caption{}
            \label{fig:fig5d}
        \end{subfigure}
    \end{subfigure}
    \caption{\textbf{Orientation column structure in the optic lobe.} (a) Three examples of orientation columns in the Dm layer, colour-coded by preferred orientation. (b) Three examples of orientation columns in the Pm layer, colour-coded by preferred orientation. (c) Quantification of tuning consistency within columns. Top: Dm layer. Bottom: Pm layer. *** indicates significance assessed via Mann-Whitney U test comparing standard deviations to baseline ($p < 10^{-20}$).}
\end{figure}
We next asked whether such tuning is preserved across the depth of the medulla. We examined the existence of column structures - clusters of neurons aligned vertically across layers that share similar orientation preferences. Putative orientation columns were defined by anchoring each analysis region around a Tm3 neuron, chosen due to their well-defined columnar morphology. For each anchor neuron, we extracted the surrounding neurons within a cylindrical region of $1.5 \times 10^4~\mathrm{nm}$ radius for both the Dm and Pm layers according to their skeleton coordinates. Each such grouping was considered as a single column. A total of 858 columns were extracted, each containing on average approximately 158.29 neurons in the Dm layers, of which 124.14 belong to Dm3 and Dm15 neuron types and 29.7 neurons in the Pm layers.

We visualised the neuron morphology by projecting a rectangular slice passing through the centre of the cylinder onto a 2D plane. This plane was defined by the two orthogonal axes perpendicular to the principal axis - the longest direction along which the neuron's morphology extends - of the Tm3 skeleton, ensuring that the projection aligned with the local column geometry. Specifically, neuron positions were transformed into this local coordinate frame by projecting each skeleton point onto the two orthogonal basis vectors to the Tm3 axis. This allowed us to flatten the 3D column into a 2D view while preserving vertical alignment across layers. We randomly selected three distinct columns each from Dm (Figure \ref{fig:fig5a}) and Pm (Figure \ref{fig:fig5b}) for visualisation. The Dm layer did exhibit orientation column organisations, with vertically clustered neurons sharing similar orientation preferences. However, Dm3 (Dm3v, Dm3p, Dm3q) and Dm15 neurons seemed to form a distinct horizontal layer that bisects each column into upper and lower segments, creating a clear visual stratification within individual columns (Figure \ref{fig:fig5a}). This pattern was observed consistently across columns, indicating a widespread structural feature in the medulla. Notably, these neuron types exhibited only a single orientation preference (Dm3v: $\approx 0\degree$; Dm3p: $\approx 60\degree$; Dm3q: $\approx 120\degree$; Dm15: $\approx 100\degree$), in contrast to other types that exhibited broader tuning curves or gradual changes in preferred orientation across space (Figure \ref{fig:fig2a}). On the other hand, the Pm layer also exhibited orientation column structure (Figure \ref{fig:fig5b}), though there are a few columns that seem less pronounced in the structure (left two panels of Figure \ref{fig:fig5b}).

To quantify this alignment, we measured the circular standard deviation (see Appendix~\ref{sec:circular_standard_deviation}) of preferred orientations within each column in the Dm layer. The circular standard deviation was computed over the preferred orientations of all voxels within the column region, providing a measure of how tightly aligned the local tuning preferences are. Lower values indicated stronger orientation coherence within the column. To further ensure that the observed orientation columns were not an artefact of the simulation framework or analysis pipeline, we generated a null baseline by randomising the preferred orientations across neurons while preserving their spatial locations and column definitions. The resulting distribution revealed that most Dm columns, excluding Dm3 and Dm15, exhibited sharp tuning, with a peak at low deviation values and a minority displaying broader orientation spread (top panel of Figure \ref{fig:fig5d}). We repeated the same analysis for columns in the Pm layer. In contrast to Dm, the Pm layer showed weaker columnar correspondence, with a distribution of standard deviations centred closer to the shuffled baseline (bottom panel of Figure \ref{fig:fig5d}). Compared to the Pm layer, the Dm layer exhibited a more pronounced separation between the observed and baseline distributions, indicating stronger tuning coherence.

Together, these results demonstrate that \textit{Drosophila's} optic lobe contains not only intra-layer orientation maps, but also orientation columns across different layers of Dm, reminiscent of columnar architectures in vertebrate visual systems.

\section{Conclusion and Discussion}
In this study, we investigated the orientation preferences in the \textit{Drosophila} visual system and explored the spatial organisation of these preferences, revealing a structured pattern. While these maps are less discretely defined than in mammals, likely due to the lower neuronal density, the presence of spatially organised orientation selectivity and columnar structure suggests that \textit{Drosophila} possesses a rudimentary form of orientation map. Biologically, the organisation shows a population-level bias toward tangential tuning, where neurons prefer orientations aligned with the local contour of the visual field and may reflect an adaptation for encoding object boundaries or motion, enhancing contrast sensitivity across the curved surface of the compound eye \cite{Borst_Euler}. 

The results align with studies in other non-mammalian species, such as birds, turtles and zebrafishes, where sharply tuned orientation-selective neurons are found in non-laminated visual pathways as mentioned earlier \cite{Li_Xiao, Fournier_Christian, Nevin_Robles, Romano_Pietri}. However, as of now, \textit{Drosophila} is currently the only invertebrate where spatially structured orientation preferences, resembling a primitive orientation map, have been computationally demonstrated. As such, we hypothesise that structured visual maps may reflect convergent evolution, arising from shared computational demands rather than shared anatomy. The \textit{Drosophila} ommatidium \(\rightarrow\) Dm \(\rightarrow\) lobula circuit may serve a functionally analogous role to the mammalian retina \(\rightarrow\) LGN \(\rightarrow\) V1 pathway \cite{Alonso_Usrey, Hansel_Vreeswijk, Kafaligonul_Breitmeyer}, suggesting that efficient visual processing follows common architectural motifs across species. Mechanistically, dendritic integration, closed-loop motifs, and lateral inhibition likely contribute to orientation tuning in \textit{Drosophila}, as proposed in prior works \cite{Angelucci_Bressloff, Jung_Almasi, Seung}. Consistent with this interpretation, analysis of synaptic connectivity by orientation preference (Appendix \ref{sec:orientation_connectivity}) revealed that neurons with similar tuning are more likely to form recurrent connections. The study by \citet{Klapoetke_2022} reveals functionally clustered representations of complex visual features in the lobular columnar neurons projecting into the central brain. In contrast, out work focuses on orientation selectivity in the earlier stages of the optic lobe, based on connectome-driven circuit structure. Together, these complementary findings highlight how structured visual maps may extend from early visual encoding to higher-order feature integration within the \textit{Drosophila} brain.


The orientation maps at the early stages of visual processing provide critical advantages for efficient visual detection for the flies. The neurons that selectively respond to similar orientations are spatially adjacent to each other, enabling them to fire maximally when a line or edge appears at their preferred orientation, which in mammals is achieved by long-distance connections among columns with similar orientation preference. The columns are arranged in a primitive pinwheel-like map, enabling more stable and precise edge perception. This organisation supports higher-level vision by feeding processed orientation data to the central brain for shape recognition and motion detection. \textit{Drosophila} serves as a powerful minimal model for exploring core principles of vision, with potential applications in bio-inspired machine vision and low-power autonomous systems.

\paragraph{Limitations.}
This study relies on biologically grounded simulations and lacks in vivo validation, warranting experimental follow-up to confirm our predictions. Similar to prior work based on anatomical reconstructions \cite{Seung}, our approach emphasised predictive modelling to inform future experiments. The use of a simplified LIF model omits nonlinear firing dynamics and neurotransmitter effects, which may affect network behaviour. However, similar simplified models have been effectively used in large-scale cortical simulations to gain insights into emergent network dynamics \cite{Shiu_Sterne, Gavenas_2024}.

\begin{ack}
This work was supported by the National Key Research and Development Program of China under Grant 2021ZD0200301, the National Natural Science Foundation of China under Grant U2341228, the Fundamental and Interdisciplinary Disciplines Breakthrough Plan of the Ministry of Education of China under Grant JYB2025XDXM504, and Ant Group under Grant 20252000099.
\end{ack}

\bibliographystyle{plainnat}
\bibliography{references} 

\clearpage
\section*{NeurIPS Paper Checklist}

\begin{enumerate}

\item {\bf Claims}
    \item[] Question: Do the main claims made in the abstract and introduction accurately reflect the paper's contributions and scope?
    \item[] Answer: \answerYes{}
    \item[] Justification: The paper's abstract and introduction clearly state that the work tests whether structured orientation maps can emerge in Drosophila purely from biological connectivity, and the results fully support this claim through connectome-driven simulations and analysis. 
    \item[] Guidelines:
    \begin{itemize}
        \item The answer NA means that the abstract and introduction do not include the claims made in the paper.
        \item The abstract and/or introduction should clearly state the claims made, including the contributions made in the paper and important assumptions and limitations. A No or NA answer to this question will not be perceived well by the reviewers. 
        \item The claims made should match theoretical and experimental results, and reflect how much the results can be expected to generalize to other settings. 
        \item It is fine to include aspirational goals as motivation as long as it is clear that these goals are not attained by the paper. 
    \end{itemize}

\item {\bf Limitations}
    \item[] Question: Does the paper discuss the limitations of the work performed by the authors?
    \item[] Answer: \answerYes{}
    \item[] Justification: The paper includes a dedicated "Limitations" section where we clearly acknowledge that our findings are based on simulations without vivo validation, and discuss the modelling simplifications. 
    \item[] Guidelines:
    \begin{itemize}
        \item The answer NA means that the paper has no limitation while the answer No means that the paper has limitations, but those are not discussed in the paper. 
        \item The authors are encouraged to create a separate "Limitations" section in their paper.
        \item The paper should point out any strong assumptions and how robust the results are to violations of these assumptions (e.g., independence assumptions, noiseless settings, model well-specification, asymptotic approximations only holding locally). The authors should reflect on how these assumptions might be violated in practice and what the implications would be.
        \item The authors should reflect on the scope of the claims made, e.g., if the approach was only tested on a few datasets or with a few runs. In general, empirical results often depend on implicit assumptions, which should be articulated.
        \item The authors should reflect on the factors that influence the performance of the approach. For example, a facial recognition algorithm may perform poorly when image resolution is low or images are taken in low lighting. Or a speech-to-text system might not be used reliably to provide closed captions for online lectures because it fails to handle technical jargon.
        \item The authors should discuss the computational efficiency of the proposed algorithms and how they scale with dataset size.
        \item If applicable, the authors should discuss possible limitations of their approach to address problems of privacy and fairness.
        \item While the authors might fear that complete honesty about limitations might be used by reviewers as grounds for rejection, a worse outcome might be that reviewers discover limitations that aren't acknowledged in the paper. The authors should use their best judgment and recognize that individual actions in favor of transparency play an important role in developing norms that preserve the integrity of the community. Reviewers will be specifically instructed to not penalize honesty concerning limitations.
    \end{itemize}

\item {\bf Theory assumptions and proofs}
    \item[] Question: For each theoretical result, does the paper provide the full set of assumptions and a complete (and correct) proof?
    \item[] Answer: \answerNA{}
    \item[] Justification: The paper does not include formal theoretical results, theorems, or proofs; it is an experimental and simulaiton-driven study grounded in biological data rather than mathematical derivations. 
    \item[] Guidelines:
    \begin{itemize}
        \item The answer NA means that the paper does not include theoretical results. 
        \item All the theorems, formulas, and proofs in the paper should be numbered and cross-referenced.
        \item All assumptions should be clearly stated or referenced in the statement of any theorems.
        \item The proofs can either appear in the main paper or the supplemental material, but if they appear in the supplemental material, the authors are encouraged to provide a short proof sketch to provide intuition. 
        \item Inversely, any informal proof provided in the core of the paper should be complemented by formal proofs provided in appendix or supplemental material.
        \item Theorems and Lemmas that the proof relies upon should be properly referenced. 
    \end{itemize}

    \item {\bf Experimental result reproducibility}
    \item[] Question: Does the paper fully disclose all the information needed to reproduce the main experimental results of the paper to the extent that it affects the main claims and/or conclusions of the paper (regardless of whether the code and data are provided or not)?
    \item[] Answer: \answerYes{}
    \item[] Justification: The paper fully discloses the experimental setup, including model parameters, connectome sources, input stimulus design, simulation details and analysis equations, providing sufficient information for others to reproduce the main results.
    \item[] Guidelines:
    \begin{itemize}
        \item The answer NA means that the paper does not include experiments.
        \item If the paper includes experiments, a No answer to this question will not be perceived well by the reviewers: Making the paper reproducible is important, regardless of whether the code and data are provided or not.
        \item If the contribution is a dataset and/or model, the authors should describe the steps taken to make their results reproducible or verifiable. 
        \item Depending on the contribution, reproducibility can be accomplished in various ways. For example, if the contribution is a novel architecture, describing the architecture fully might suffice, or if the contribution is a specific model and empirical evaluation, it may be necessary to either make it possible for others to replicate the model with the same dataset, or provide access to the model. In general. releasing code and data is often one good way to accomplish this, but reproducibility can also be provided via detailed instructions for how to replicate the results, access to a hosted model (e.g., in the case of a large language model), releasing of a model checkpoint, or other means that are appropriate to the research performed.
        \item While NeurIPS does not require releasing code, the conference does require all submissions to provide some reasonable avenue for reproducibility, which may depend on the nature of the contribution. For example
        \begin{enumerate}
            \item If the contribution is primarily a new algorithm, the paper should make it clear how to reproduce that algorithm.
            \item If the contribution is primarily a new model architecture, the paper should describe the architecture clearly and fully.
            \item If the contribution is a new model (e.g., a large language model), then there should either be a way to access this model for reproducing the results or a way to reproduce the model (e.g., with an open-source dataset or instructions for how to construct the dataset).
            \item We recognize that reproducibility may be tricky in some cases, in which case authors are welcome to describe the particular way they provide for reproducibility. In the case of closed-source models, it may be that access to the model is limited in some way (e.g., to registered users), but it should be possible for other researchers to have some path to reproducing or verifying the results.
        \end{enumerate}
    \end{itemize}

\item {\bf Open access to data and code}
    \item[] Question: Does the paper provide open access to the data and code, with sufficient instructions to faithfully reproduce the main experimental results, as described in supplemental material?
    \item[] Answer: \answerYes{} 
    \item[] Justification: The full code for simulation and analysis is appended in the Github link in the Appendix \ref{sec:data-availability}, along with sufficient instructions to reproduce the main experimental results.
    \item[] Guidelines:
    \begin{itemize}
        \item The answer NA means that paper does not include experiments requiring code.
        \item Please see the NeurIPS code and data submission guidelines (\url{https://nips.cc/public/guides/CodeSubmissionPolicy}) for more details.
        \item While we encourage the release of code and data, we understand that this might not be possible, so “No” is an acceptable answer. Papers cannot be rejected simply for not including code, unless this is central to the contribution (e.g., for a new open-source benchmark).
        \item The instructions should contain the exact command and environment needed to run to reproduce the results. See the NeurIPS code and data submission guidelines (\url{https://nips.cc/public/guides/CodeSubmissionPolicy}) for more details.
        \item The authors should provide instructions on data access and preparation, including how to access the raw data, preprocessed data, intermediate data, and generated data, etc.
        \item The authors should provide scripts to reproduce all experimental results for the new proposed method and baselines. If only a subset of experiments are reproducible, they should state which ones are omitted from the script and why.
        \item At submission time, to preserve anonymity, the authors should release anonymized versions (if applicable).
        \item Providing as much information as possible in supplemental material (appended to the paper) is recommended, but including URLs to data and code is permitted.
    \end{itemize}

\item {\bf Experimental setting/details}
    \item[] Question: Does the paper specify all the training and test details (e.g., data splits, hyperparameters, how they were chosen, type of optimizer, etc.) necessary to understand the results?
    \item[] Answer: \answerYes{}
    \item[] Justification: The paper specifies all necessary experimental details, including the LIF model parameters, input stimulus construction, simulation setup and neuron types studied. 
    \item[] Guidelines:
    \begin{itemize}
        \item The answer NA means that the paper does not include experiments.
        \item The experimental setting should be presented in the core of the paper to a level of detail that is necessary to appreciate the results and make sense of them.
        \item The full details can be provided either with the code, in appendix, or as supplemental material.
    \end{itemize}

\item {\bf Experiment statistical significance}
    \item[] Question: Does the paper report error bars suitably and correctly defined or other appropriate information about the statistical significance of the experiments?
    \item[] Answer: \answerYes{} 
    \item[] Justification: The paper reports statistical robustness through visual summaries and null-model comparisons. Specifically, distributions of orientation tuning sharpness are shown using histograms, and baseline significance is assessed via shuffled control data. While traditional error bars are not uniformly presented, the use of population-wide distribution and comparison to randomised baselines appropriately conveys statistical reliability for the claims made.
    \item[] Guidelines:
    \begin{itemize}
        \item The answer NA means that the paper does not include experiments.
        \item The authors should answer "Yes" if the results are accompanied by error bars, confidence intervals, or statistical significance tests, at least for the experiments that support the main claims of the paper.
        \item The factors of variability that the error bars are capturing should be clearly stated (for example, train/test split, initialization, random drawing of some parameter, or overall run with given experimental conditions).
        \item The method for calculating the error bars should be explained (closed form formula, call to a library function, bootstrap, etc.)
        \item The assumptions made should be given (e.g., Normally distributed errors).
        \item It should be clear whether the error bar is the standard deviation or the standard error of the mean.
        \item It is OK to report 1-sigma error bars, but one should state it. The authors should preferably report a 2-sigma error bar than state that they have a 96\% CI, if the hypothesis of Normality of errors is not verified.
        \item For asymmetric distributions, the authors should be careful not to show in tables or figures symmetric error bars that would yield results that are out of range (e.g. negative error rates).
        \item If error bars are reported in tables or plots, The authors should explain in the text how they were calculated and reference the corresponding figures or tables in the text.
    \end{itemize}

\item {\bf Experiments compute resources}
    \item[] Question: For each experiment, does the paper provide sufficient information on the computer resources (type of compute workers, memory, time of execution) needed to reproduce the experiments?
    \item[] Answer: \answerYes{}
    \item[] Justification: The paper specifies the computing resources used by the simulation in Appendix \ref{sec:data-availability}. 
    \item[] Guidelines:
    \begin{itemize}
        \item The answer NA means that the paper does not include experiments.
        \item The paper should indicate the type of compute workers CPU or GPU, internal cluster, or cloud provider, including relevant memory and storage.
        \item The paper should provide the amount of compute required for each of the individual experimental runs as well as estimate the total compute. 
        \item The paper should disclose whether the full research project required more compute than the experiments reported in the paper (e.g., preliminary or failed experiments that didn't make it into the paper). 
    \end{itemize}
    
\item {\bf Code of ethics}
    \item[] Question: Does the research conducted in the paper conform, in every respect, with the NeurIPS Code of Ethics \url{https://neurips.cc/public/EthicsGuidelines}?
    \item[] Answer: \answerYes{}
    \item[] Justification: The research fully conforms to the NeurIPS Code of Ethics, focusing on non-invasive, simulation-based studies using publicly available biological datasets, with no involvement of human subjects, personal data, or potentially harmful applications.
    \item[] Guidelines:
    \begin{itemize}
        \item The answer NA means that the authors have not reviewed the NeurIPS Code of Ethics.
        \item If the authors answer No, they should explain the special circumstances that require a deviation from the Code of Ethics.
        \item The authors should make sure to preserve anonymity (e.g., if there is a special consideration due to laws or regulations in their jurisdiction).
    \end{itemize}

\item {\bf Broader impacts}
    \item[] Question: Does the paper discuss both potential positive societal impacts and negative societal impacts of the work performed?
    \item[] Answer: \answerYes{}.
    \item[] Justification: The paper discusses the broader impacts of its findings by suggesting that understanding minimalistic biological circuits could inspire lightweight, sustainable AI systems.
    \item[] Guidelines:
    \begin{itemize}
        \item The answer NA means that there is no societal impact of the work performed.
        \item If the authors answer NA or No, they should explain why their work has no societal impact or why the paper does not address societal impact.
        \item Examples of negative societal impacts include potential malicious or unintended uses (e.g., disinformation, generating fake profiles, surveillance), fairness considerations (e.g., deployment of technologies that could make decisions that unfairly impact specific groups), privacy considerations, and security considerations.
        \item The conference expects that many papers will be foundational research and not tied to particular applications, let alone deployments. However, if there is a direct path to any negative applications, the authors should point it out. For example, it is legitimate to point out that an improvement in the quality of generative models could be used to generate deepfakes for disinformation. On the other hand, it is not needed to point out that a generic algorithm for optimizing neural networks could enable people to train models that generate Deepfakes faster.
        \item The authors should consider possible harms that could arise when the technology is being used as intended and functioning correctly, harms that could arise when the technology is being used as intended but gives incorrect results, and harms following from (intentional or unintentional) misuse of the technology.
        \item If there are negative societal impacts, the authors could also discuss possible mitigation strategies (e.g., gated release of models, providing defenses in addition to attacks, mechanisms for monitoring misuse, mechanisms to monitor how a system learns from feedback over time, improving the efficiency and accessibility of ML).
    \end{itemize}
    
\item {\bf Safeguards}
    \item[] Question: Does the paper describe safeguards that have been put in place for responsible release of data or models that have a high risk for misuse (e.g., pretrained language models, image generators, or scraped datasets)?
    \item[] Answer: \answerNA{}
    \item[] Justification: The paper does not release any pretrained models, datasets, or tools that pose a risk of misuse or dual-use concerns, as it focuses solely on simulated neural network analyses based on publicly available biological data.
    \item[] Guidelines:
    \begin{itemize}
        \item The answer NA means that the paper poses no such risks.
        \item Released models that have a high risk for misuse or dual-use should be released with necessary safeguards to allow for controlled use of the model, for example by requiring that users adhere to usage guidelines or restrictions to access the model or implementing safety filters. 
        \item Datasets that have been scraped from the Internet could pose safety risks. The authors should describe how they avoided releasing unsafe images.
        \item We recognize that providing effective safeguards is challenging, and many papers do not require this, but we encourage authors to take this into account and make a best faith effort.
    \end{itemize}

\item {\bf Licenses for existing assets}
    \item[] Question: Are the creators or original owners of assets (e.g., code, data, models), used in the paper, properly credited and are the license and terms of use explicitly mentioned and properly respected?
    \item[] Answer: \answerYes{}
    \item[] Justification: The paper properly cites the FAFB connectome and other publicly available Drosophila datasets, referencing the original sources. 
    \item[] Guidelines:
    \begin{itemize}
        \item The answer NA means that the paper does not use existing assets.
        \item The authors should cite the original paper that produced the code package or dataset.
        \item The authors should state which version of the asset is used and, if possible, include a URL.
        \item The name of the license (e.g., CC-BY 4.0) should be included for each asset.
        \item For scraped data from a particular source (e.g., website), the copyright and terms of service of that source should be provided.
        \item If assets are released, the license, copyright information, and terms of use in the package should be provided. For popular datasets, \url{paperswithcode.com/datasets} has curated licenses for some datasets. Their licensing guide can help determine the license of a dataset.
        \item For existing datasets that are re-packaged, both the original license and the license of the derived asset (if it has changed) should be provided.
        \item If this information is not available online, the authors are encouraged to reach out to the asset's creators.
    \end{itemize}

\item {\bf New assets}
    \item[] Question: Are new assets introduced in the paper well documented and is the documentation provided alongside the assets?
    \item[] Answer: \answerNA{}
    \item[] Justification: The paper does not introduce new datasets or models beyond the use of existing publicly available connectome data and simulated networks. 
    \item[] Guidelines:
    \begin{itemize}
        \item The answer NA means that the paper does not release new assets.
        \item Researchers should communicate the details of the dataset/code/model as part of their submissions via structured templates. This includes details about training, license, limitations, etc. 
        \item The paper should discuss whether and how consent was obtained from people whose asset is used.
        \item At submission time, remember to anonymize your assets (if applicable). You can either create an anonymized URL or include an anonymized zip file.
    \end{itemize}

\item {\bf Crowdsourcing and research with human subjects}
    \item[] Question: For crowdsourcing experiments and research with human subjects, does the paper include the full text of instructions given to participants and screenshots, if applicable, as well as details about compensation (if any)? 
    \item[] Answer: \answerNA{}
    \item[] Justification: The study does not involve any crowdsourcing, surveys, or human subject research, as all experiments were performed using simulated neural models based on biological datasets. 
    \item[] Guidelines:
    \begin{itemize}
        \item The answer NA means that the paper does not involve crowdsourcing nor research with human subjects.
        \item Including this information in the supplemental material is fine, but if the main contribution of the paper involves human subjects, then as much detail as possible should be included in the main paper. 
        \item According to the NeurIPS Code of Ethics, workers involved in data collection, curation, or other labor should be paid at least the minimum wage in the country of the data collector. 
    \end{itemize}

\item {\bf Institutional review board (IRB) approvals or equivalent for research with human subjects}
    \item[] Question: Does the paper describe potential risks incurred by study participants, whether such risks were disclosed to the subjects, and whether Institutional Review Board (IRB) approvals (or an equivalent approval/review based on the requirements of your country or institution) were obtained?
    \item[] Answer: \answerNA{}
    \item[] Justification: The paper does not involve human subjects, crowdsourcing, or participant-based studies, and therefore IRB approvals is not applicable. 
    \item[] Guidelines:
    \begin{itemize}
        \item The answer NA means that the paper does not involve crowdsourcing nor research with human subjects.
        \item Depending on the country in which research is conducted, IRB approval (or equivalent) may be required for any human subjects research. If you obtained IRB approval, you should clearly state this in the paper. 
        \item We recognize that the procedures for this may vary significantly between institutions and locations, and we expect authors to adhere to the NeurIPS Code of Ethics and the guidelines for their institution. 
        \item For initial submissions, do not include any information that would break anonymity (if applicable), such as the institution conducting the review.
    \end{itemize}

\item {\bf Declaration of LLM usage}
    \item[] Question: Does the paper describe the usage of LLMs if it is an important, original, or non-standard component of the core methods in this research? Note that if the LLM is used only for writing, editing, or formatting purposes and does not impact the core methodology, scientific rigorousness, or originality of the research, declaration is not required.
    \item[] Answer: \answerNA{}
    \item[] Justification: The paper's core research methods do not involve large language models (LLMs) as an important, original, or non-standard component. 
    \item[] Guidelines:
    \begin{itemize}
        \item The answer NA means that the core method development in this research does not involve LLMs as any important, original, or non-standard components.
        \item Please refer to our LLM policy (\url{https://neurips.cc/Conferences/2025/LLM}) for what should or should not be described.
    \end{itemize}

\end{enumerate}

\newpage
\appendix
\setcounter{table}{0}
\renewcommand{\thetable}{S\arabic{table}}%
\setcounter{figure}{0}
\renewcommand{\thefigure}{S\arabic{figure}}%
\setcounter{section}{0}
\section*{\Large Supplementary Material}
\section{Code and data availability}
\label{sec:data-availability}
The full code and data can be found at \url{https://github.com/JNLiew/flylif_orientation_maps}. The model ran on an Intel(R) CPU at 2.90GHz machine using 20 threads in parallel. Each thread required approximately 72 hours to complete a single round of simulation of different angles in the range 0\(\degree\) - 180\(\degree\) with an interval of \(0.4 \degree\).

\section{Experimental and analysis definition}
\subsection{LIF model design}
\label{sec:appendix-lif-design}
We implemented a conductance-based leaky integrate-and-fire (LIF) model~\cite{Churgin_Lavrentovich, Huang_Wang, Anna_Khalili, Kakaria_Bivort, Lazar_Liu, Shiu_Sterne} and integrated the FAFB dataset into this LIF model to run our simulations~\cite{Buhmann_Sheridan, Dorkenwald_Matsliah, Eckstein_Bates, Heinrich_Funke-key, Lin_Yang, codex, Mastsliah_Yu, Schlegel_Yin, Zheng_Lauritzen}. The membrane potential \( v_i \) of neuron $i$ evolves as:
\begin{align}
    \frac{dv_i}{dt} &= \frac{g_i - (v_i - V_\text{resting})}{T_\text{mbr}}, \label{equation:1}\\
    \frac{dg_i}{dt} &= -\frac{g_i}{\tau}, \label{equation:2}\\
    g_i &\leftarrow g_i + (w_{j,i} * w_{syn}) \quad \text{upon spike from neuron } j.
\end{align}
A spike is emitted when \( v_i \geq V_\text{threshold} \), after which the membrane potential is reset to \( V_\text{reset} \) for a refractory period \( T_\text{refractory} \). Synaptic transmission is delayed by a fixed latency \( T_\text{dly} \). The model used the following parameter values obtained from \textit{Drosophila} modelling or electrophysiology efforts \cite{Churgin_Lavrentovich, Anna_Khalili, Kakaria_Bivort, Paul_Pauli, Shiu_Sterne}:
\begin{itemize}
    \item \( V_\text{resting} = -52~\text{mV} \): resting potential~\cite{Kakaria_Bivort, Shiu_Sterne}
    \item \( V_\text{reset} = -52~\text{mV} \): reset potential~\cite{Kakaria_Bivort, Shiu_Sterne}
    \item \( V_\text{threshold} = -45~\text{mV} \): spiking threshold~\cite{Kakaria_Bivort, Shiu_Sterne}
    \item \( R_\text{mbr} = 10~\text{k}\Omega\,\text{cm}^2 \): membrane resistance~\cite{Kakaria_Bivort, Shiu_Sterne}
    \item \( C_\text{mbr} = 2~\mu\text{F}\,\text{cm}^{-2} \): membrane capacitance~\cite{Kakaria_Bivort, Shiu_Sterne}
    \item \( T_\text{mbr} = R_\text{mbr} \cdot C_\text{mbr} \): membrane time constant~\cite{Shiu_Sterne}
    \item \( T_\text{refractory} = 2.2~\text{ms} \): refractory period~\cite{Kakaria_Bivort, Lazar_Liu, Shiu_Sterne}
    \item \( \tau = 5~\text{ms} \): synaptic decay time constant~\cite{Anna_Khalili, Shiu_Sterne}
    \item \( T_\text{dly} = 1.8~\text{ms} \):  synaptic transmission delay~\cite{Shiu_Sterne}
    \item \( w_\text{syn} = 1.5~\text{mV} \): synaptic weight; free parameter
    \item \(g_i \): the synaptic conductance resulting from the aggregate firing of neurons presynaptic to neuron $i$.
\end{itemize}


 \( w_\text{syn}\) is a free parameter representing the strength of excitatory and inhibitory postsynaptic potentials. The value was chosen to ensure sufficient firing activity across the medulla for analysis, given the low baseline firing rates of visual neurons. The number of synapses of an upstream neuron $j$ on a downstream neuron $i$ is represented by the connectivity $w_{j,i}$ such that if neuron $j$ fires, the membrane potential of neuron $i$ changes in proportion to its connectivity $w_{j,i} \cdot w_{syn}$.

\begin{figure}[tp]
    \centering
    \begin{subfigure}[b]{0.9\linewidth} 
        \centering
        \includegraphics[width=\textwidth]{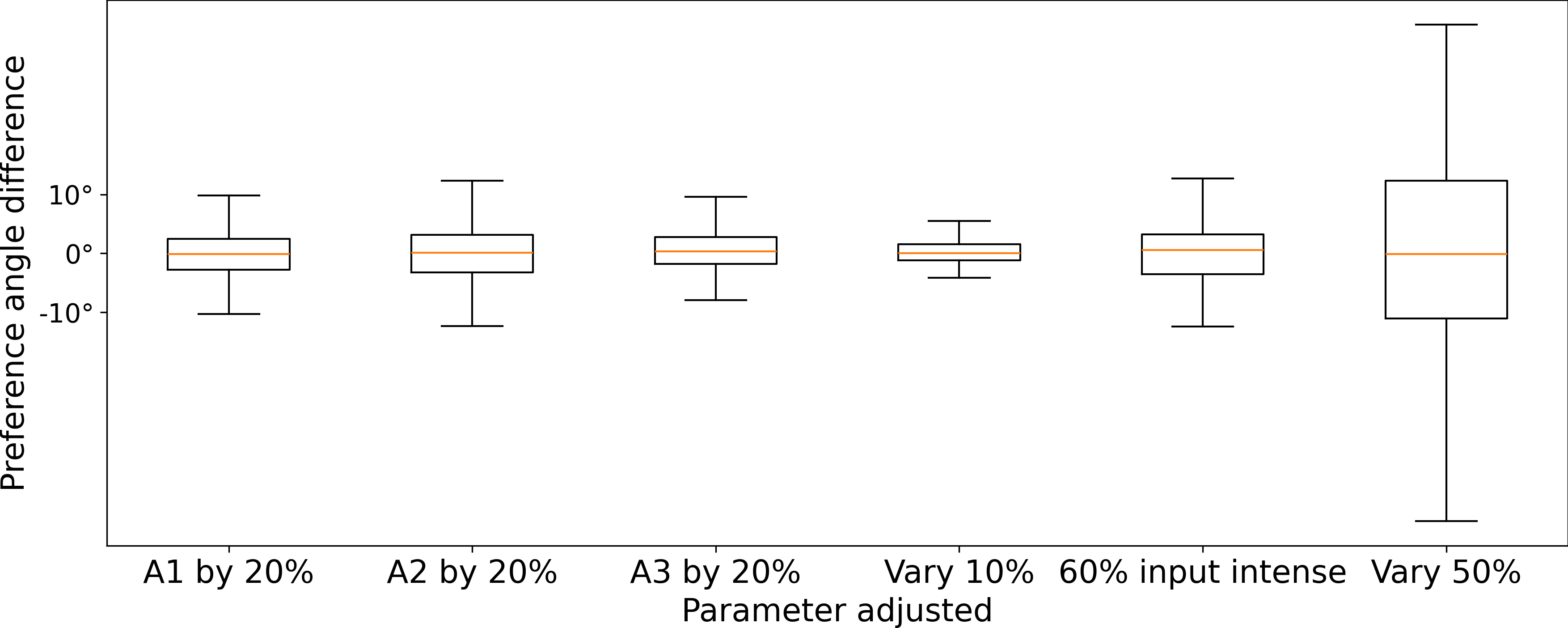}
        \caption{}
        \label{fig:para_box}
    \end{subfigure}
    \\
    \begin{subfigure}[b]{0.48\linewidth}
        \centering
        \includegraphics[width=\textwidth]{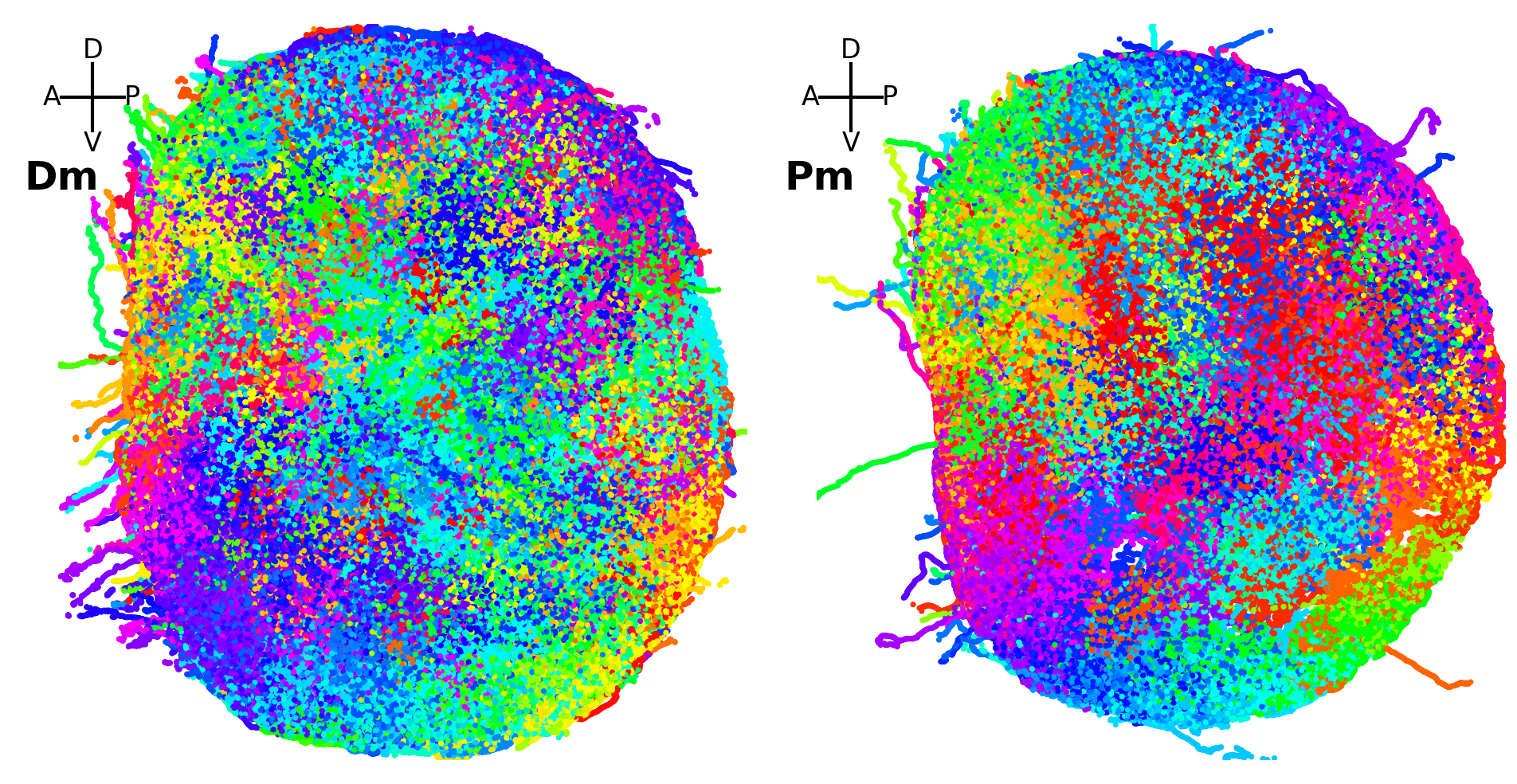}
        \caption{}
        \label{fig:para_vary_10}
    \end{subfigure} 
    \hfill
    \begin{subfigure}[b]{0.48\linewidth}
        \centering
        \includegraphics[width=\textwidth]{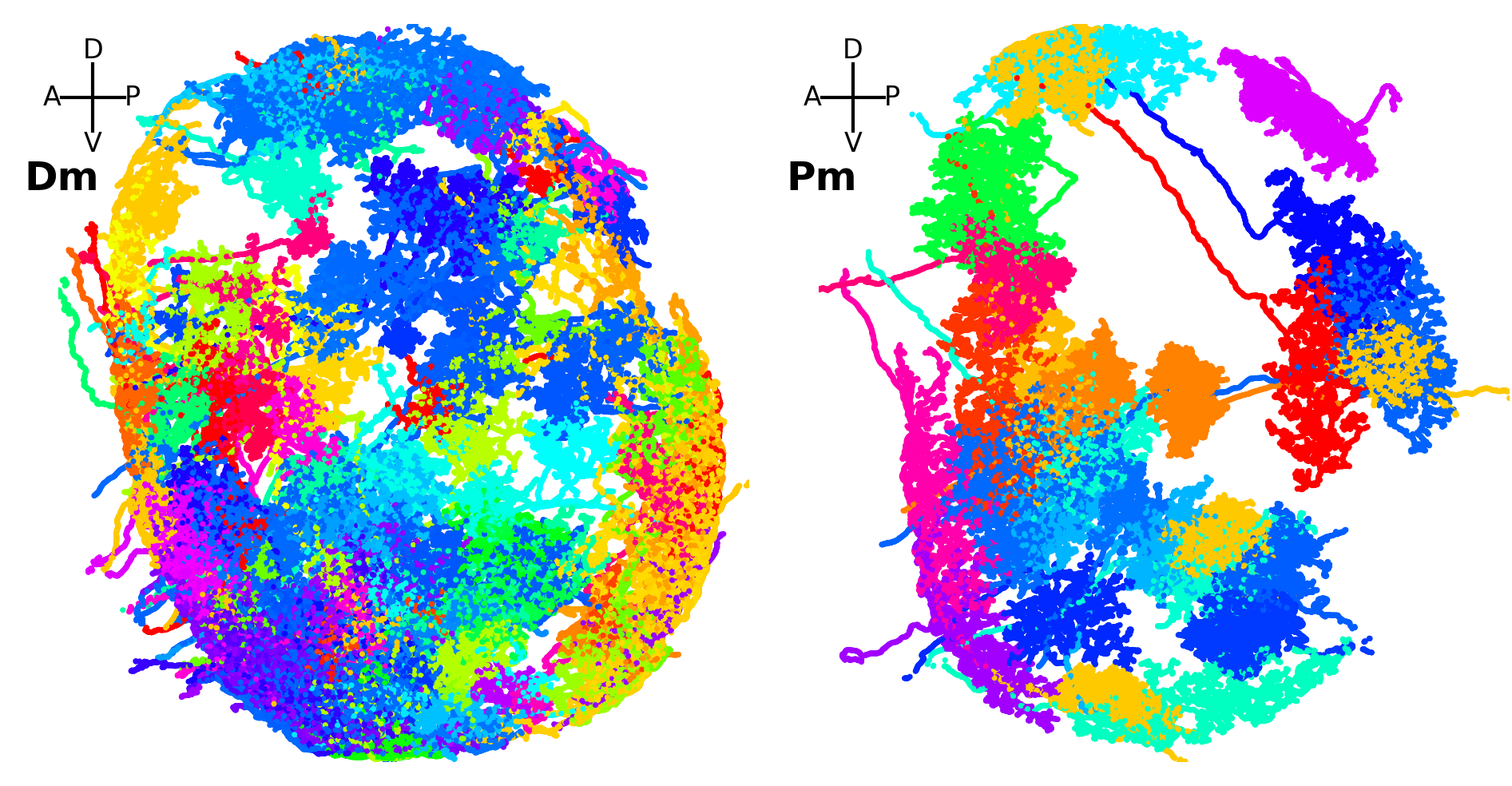}
        \caption{}
        \label{fig:para_vary_50}
    \end{subfigure} \\
    \caption{\textbf{LIF model parameter validation.} (a) Shows the difference in neurons' preferred orientation angles when stimulus parameters (A1, A2, A3) are increased by 20\% respectively. "Vary 20\%" indicates simulations where all LIF parameters (except $V_{rest}$ and $V_{threshold}$) were randomly perturbed by $\pm 10\%$. "60\% input intense" refers to simulations where all inputs were reduced to 60\%. "Vary 50\%" is similar to "Vary 10\%" but instead is randomly perturbed by $\pm 50\%$. (b) Orientation maps under 10\% LIF parameters variation. (c) Orientation maps under 50\% LIF parameters variation.}
    \label{fig:lif_parameter}
\end{figure}

To assess sensitivity to parameter perturbations, we performed a variation analysis where all parameters (except $V_{rest}$ and $V_{threshold}$) were randomly perturbed up to a fixed maximum percentage (Figure \ref{fig:para_box}). At 10\% variation, we observed minimal changes in preferred orientation across neurons (Figure \ref{fig:para_vary_10}). At 50\% orientation selectivity was noticeably degraded (Figure \ref{fig:para_vary_50}; however, surprisingly, the ODEs (Equations \ref{equation:1} and \ref{equation:2}) remained stable and did not diverge. This suggests that the system retains a degree of functional robustness even under large perturbations. 

\subsection{Stimulus designs and validation}
\label{sec:appendix-stimulus}

\begin{figure}[tp]
    \centering
    \begin{subfigure}[b]{0.48\linewidth} 
        \centering
        \includegraphics[width=\textwidth]{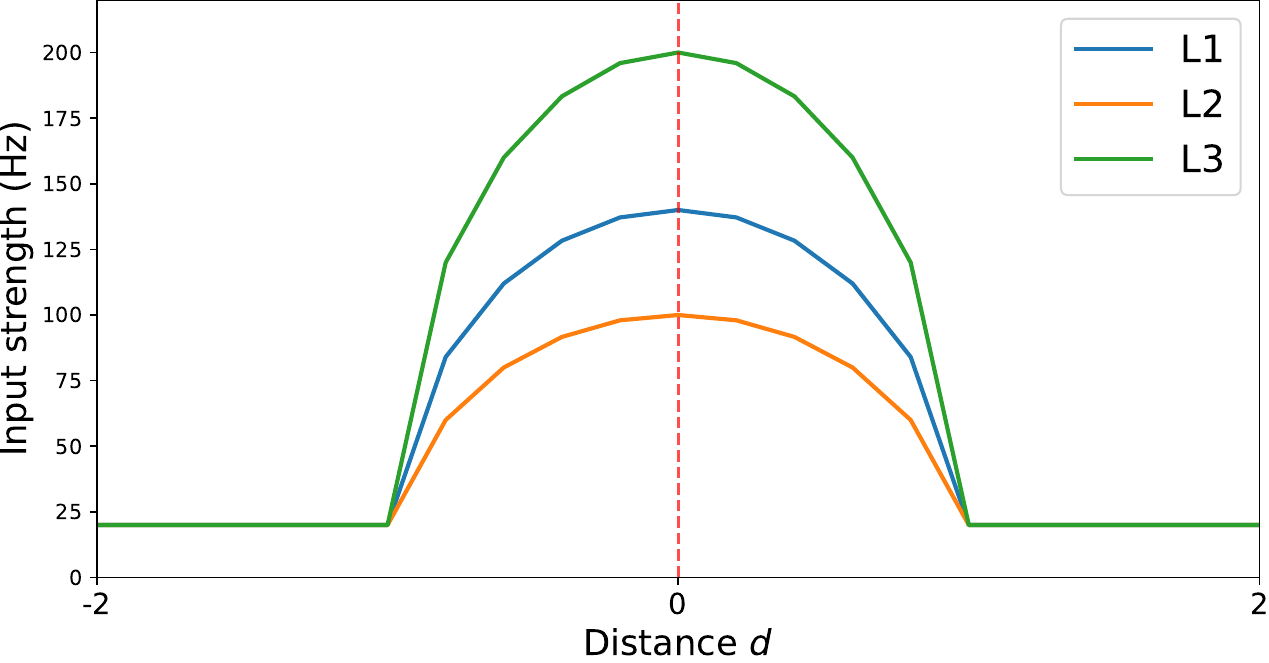}
        \caption{}
        \label{fig:ex_fig1a}
    \end{subfigure}
    \hfill
    \begin{subfigure}[b]{0.48\linewidth}
        \centering
        \includegraphics[width=\textwidth]{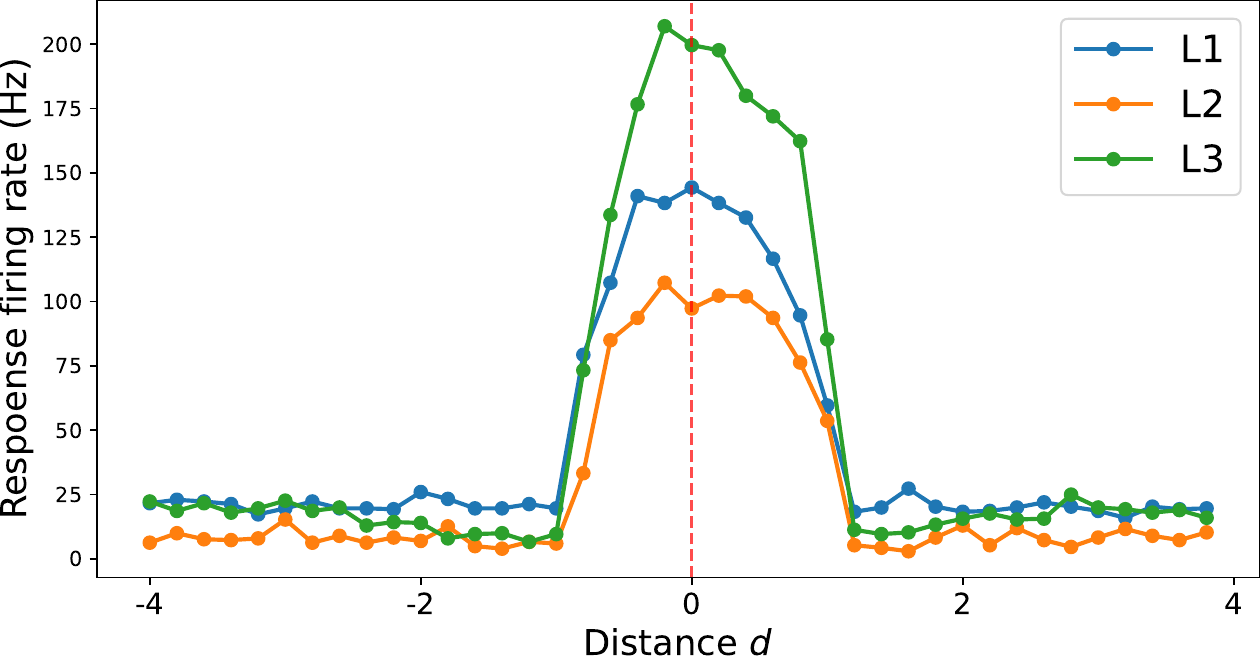}
        \caption{}
        \label{fig:ex_fig1b}
    \end{subfigure}
    \begin{subfigure}[b]{0.48\linewidth}
        \centering
        \includegraphics[width=\textwidth]{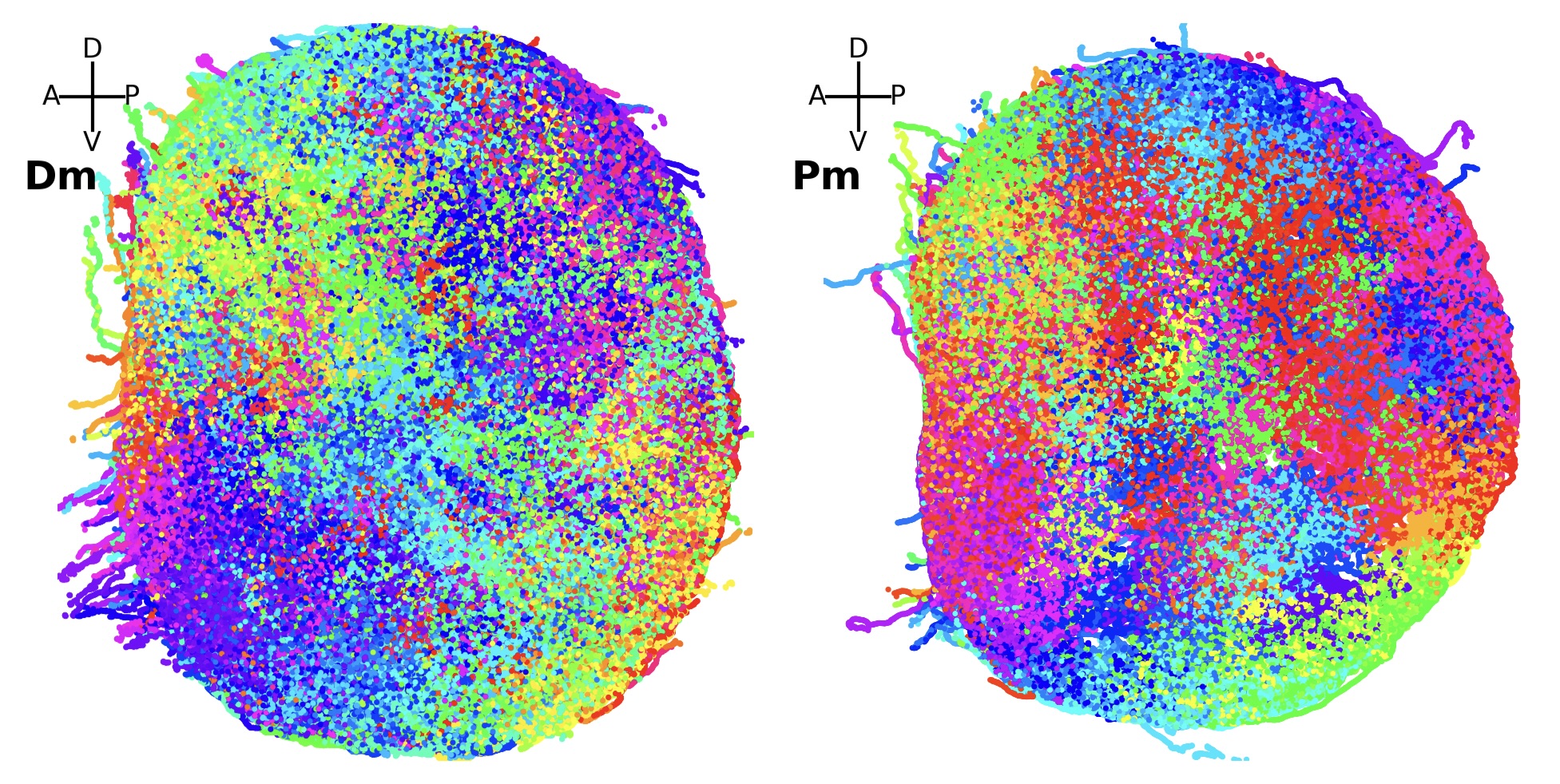}
        \caption{}
        \label{fig:para_A1}
    \end{subfigure}
    \begin{subfigure}[b]{0.48\linewidth}
        \centering
        \includegraphics[width=\textwidth]{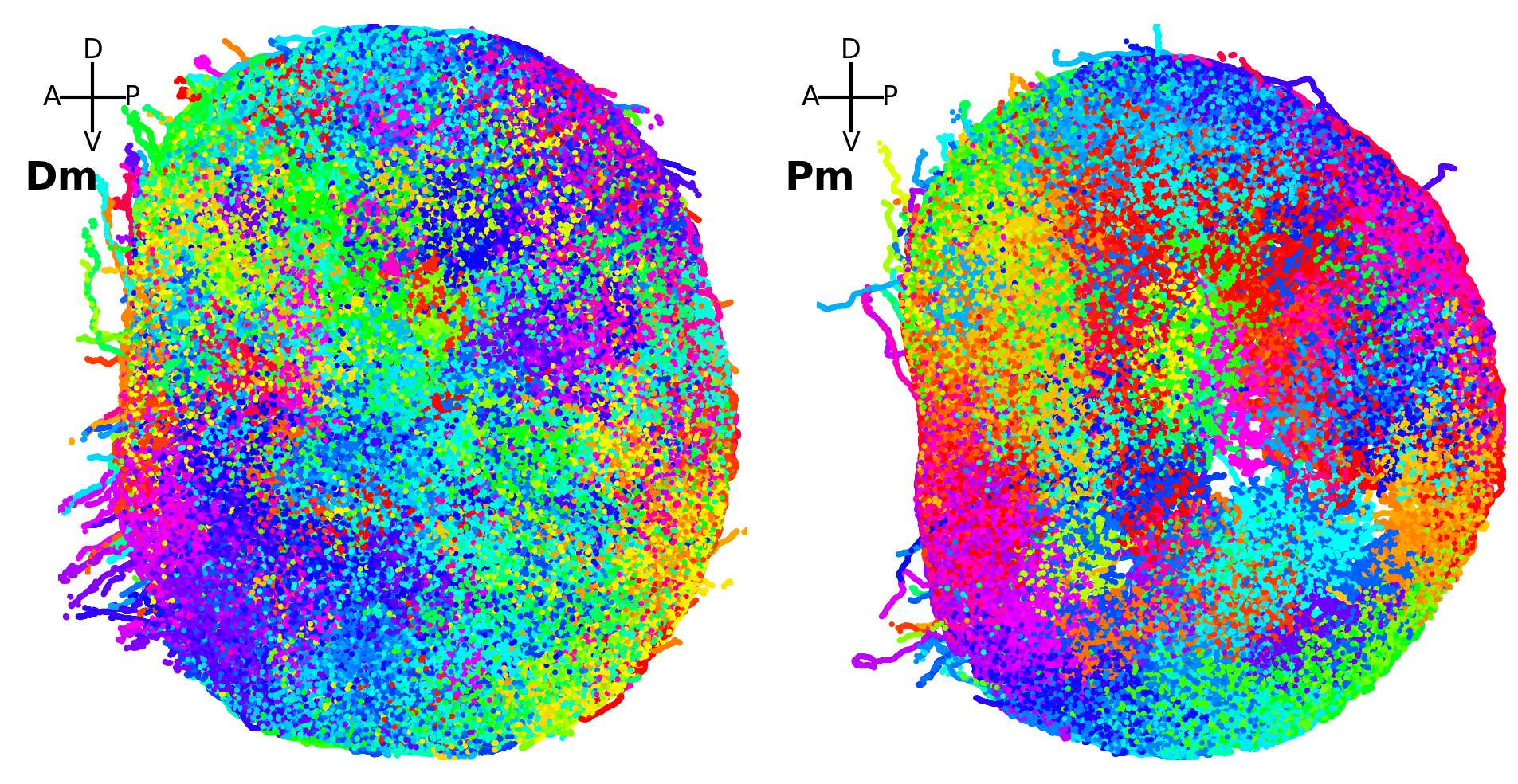}
        \caption{}
        \label{fig:para_A2}
    \end{subfigure}
    \begin{subfigure}[b]{0.48\linewidth}
        \centering
        \includegraphics[width=\textwidth]{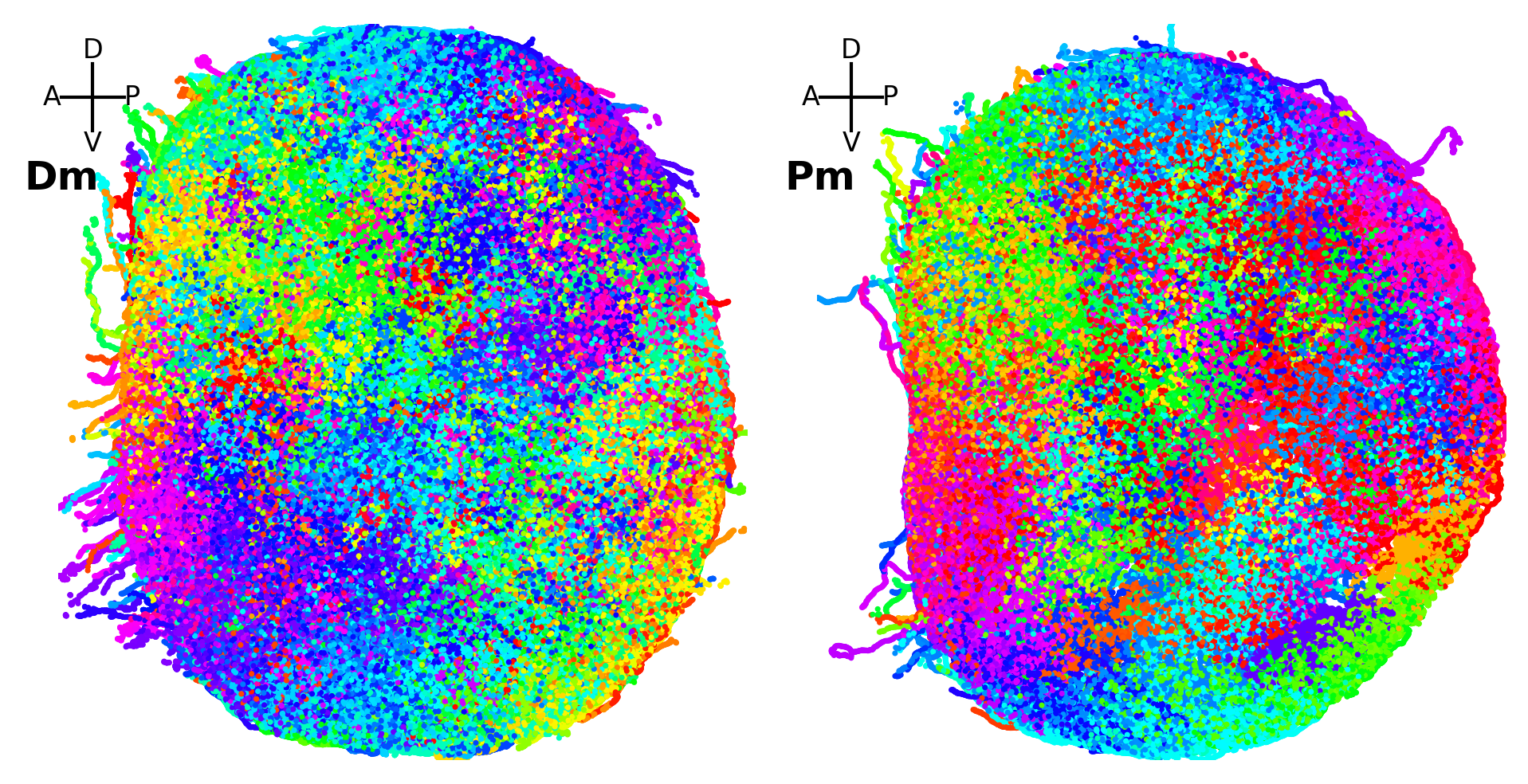}
        \caption{}
        \label{fig:para_A3}
    \end{subfigure}
    \begin{subfigure}[b]{0.3\linewidth}
        \centering
        \includegraphics[width=\textwidth]{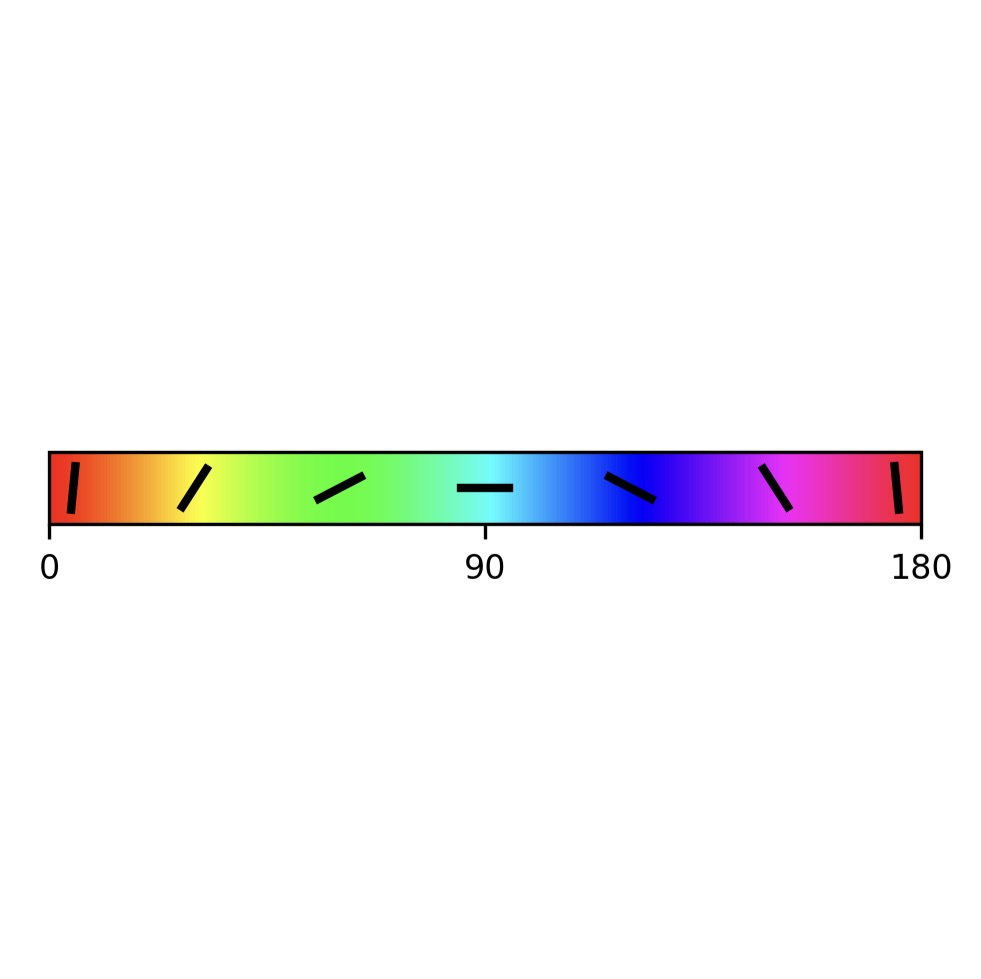}
    \end{subfigure}
    \\
    \begin{subfigure}[b]{0.48\linewidth}
        \centering
        \includegraphics[width=\textwidth]{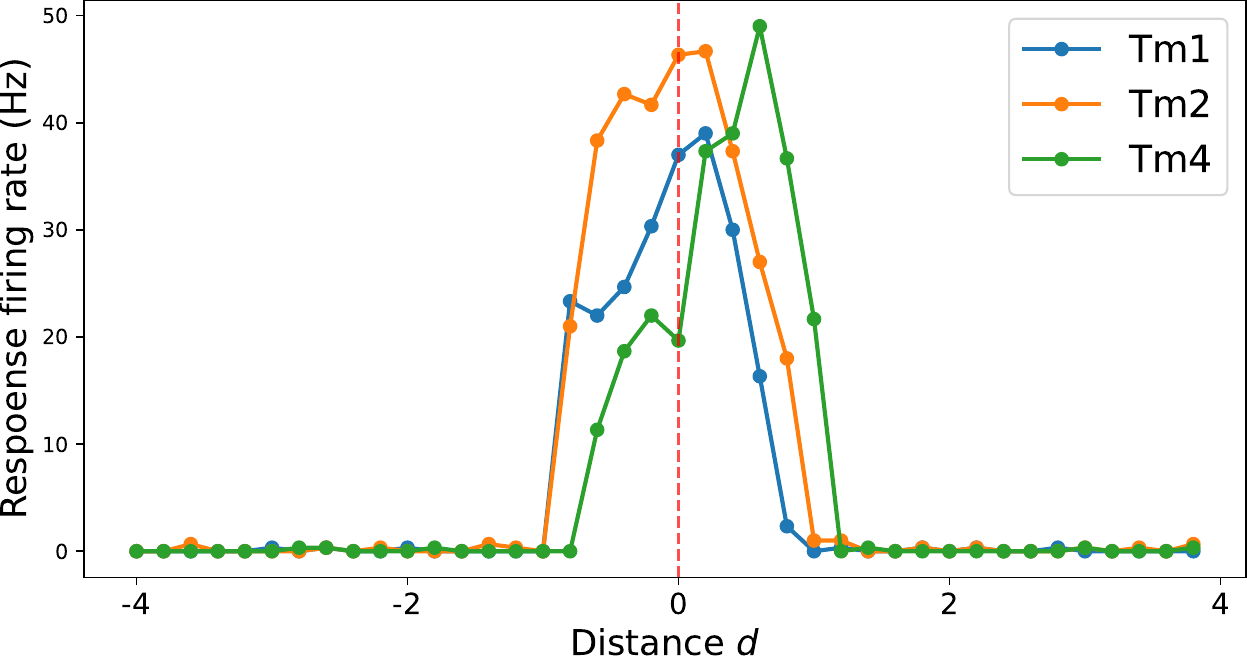}
        \caption{}
        \label{fig:ex_fig1c}
    \end{subfigure}
    \begin{subfigure}[b]{0.48\linewidth}
        \centering
        \includegraphics[width=\textwidth]{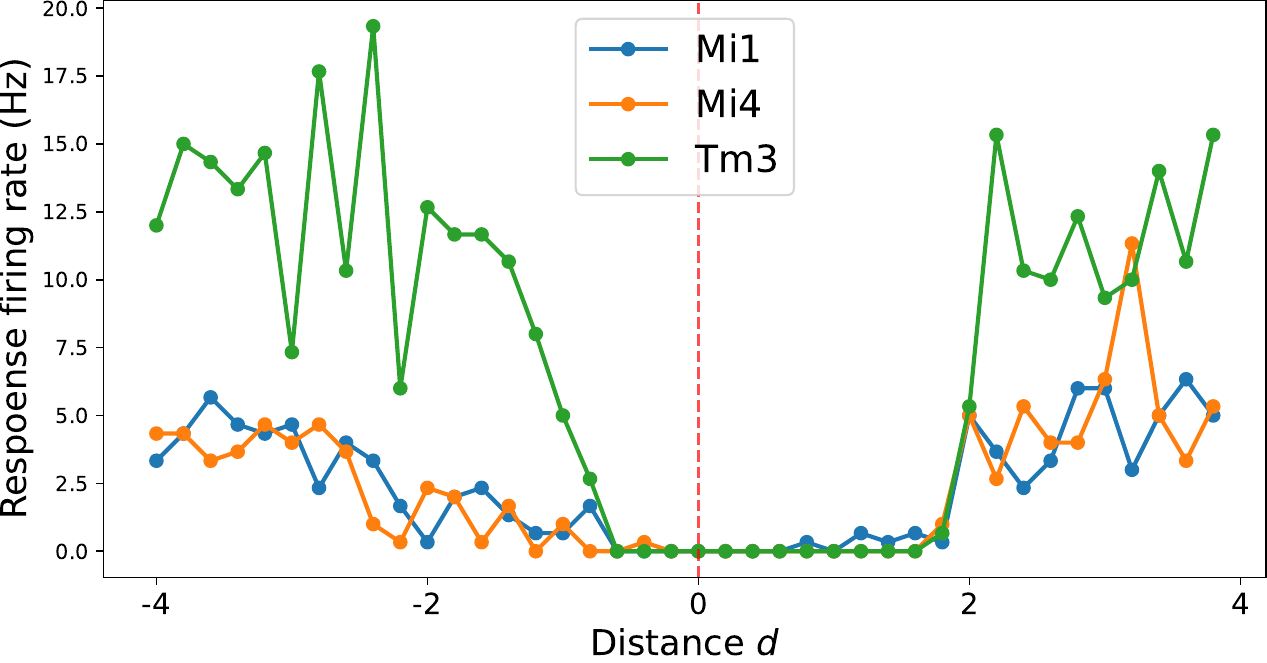}
        \caption{}
        \label{fig:ex_fig1d}
    \end{subfigure}
    \caption{\textbf{Activation of neuron types to bar-like stimuli.} (a) Modelled input strength profiles for lamina neurons L1-L3 as a function of distance $d$ from the OFF-bar stimulus. (b) Firing rate traces of a representative neuron from each of the L1-L3 types against their distance from the OFF-bar. The red line indicates where the stimulus OFF-bar is located. (c) Orientation maps of where A1 parameters are varied by 10\%. (d) Orientation maps of where A2 parameters are varied by 10\%. (e) Orientation maps of where A3 parameters are varied by 10\%. (f) Firing responses of a representative downstream OFF-response neuron from each of the Tm1, Tm2 and Tm4 types against their distance from the OFF-bar. In both (f) and (g), neurons were randomly sampled to verify that OFF-bar stimuli reliably activated these neuron types as expected. (g) Firing responses of a representative downstream ON-response neuron from each of the Mi1, Mi4 and Tm3 types against their distance from the OFF-bar. OFF-bar stimuli reliably suppress firing rates in these neurons within OFF-bar regions.}
    \label{fig:ex_fig1}
\end{figure}

To simulate early visual input, we designed a synthetic stimulus that mimics the spatial layout and angular sampling properties of \textit{Drosophila} photoreceptors. This allowed us to approximate the input that L1-L3 would receive in vivo. Our model design was guided by calcium imaging data reported in \citet{Ketkar_Burak}, which showed that L3 neurons exhibit the strongest response variation across changes in light intensity, followed by L1, with L2 showing the weakest modulation. These responses increased nonlinearly with light intensity, starting slowly and then rising more steeply. To capture this, we defined each neuron's firing rate as a function of spatial distance from a bar stimulus, which we refer to as the OFF-bar.

We assumed that spatial distance from the stimulus approximates variations in local light intensity across the visual field. The firing rate of neuron group L1-L3 was then defined as:
\begin{equation}\label{eq:1}
    \text{FiringRate}_{Lx}(d) = 
    \begin{cases}
        A_x \cdot \sqrt{1 - d^2} \cdot \frac{\max(fr)}{10}, & \text{if } |d| < 1, \\
        B_x, & \text{otherwise},
    \end{cases}
\end{equation}
where $d$ denotes the distance from the centre of the ommatidium to the OFF-bar in the visual field, measured in units where $d=1$ corresponds to the distance between the centres of two neighbouring ommatidia. $\max(fr)$ represents the maximum firing rate, set to $200$Hz. The parameters $A_x$ and $B_x$, layer-specific gain and baseline levels, were estimated by fitting model responses to experimental data \cite{Ketkar_Burak}:
\begin{itemize}
    \item L1: \( A_1 = 7 \), \(B_1 = 20 \) Hz
    \item L2: \( A_2 = 5 \), \(B_2 = 20 \) Hz
    \item L3: \(A_3 = 10 \), \(B_3 = 20 \) Hz.
\end{itemize}
The configuration ($A_3 > A_1 > A_2$) captured the relative sensitivity of L1-L3, as shown in Figure \ref{fig:ex_fig1a}. The baseline parameter $B_x$ accounted for residual activity even when the calcium signal is low or flat, as $\Delta F/F = 0$ does not necessarily imply zero firing due to background fluorescence and imaging noise. We therefore set $B_x= 20$Hz to reflect plausible baseline activity. To assess the robustness of our results to the gain parameters $A_x$, we conducted a sensitivity analysis by decreasing each of $A_1$, $A_2$ and $A_3$ by $20\%$, one at a time. The changes in well-fit downstream neurons' preferred orientation (measured by preference angle difference) remained small across all conditions (Figure \ref{fig:para_box}), and the qualitative tuning properties were preserved. Similarly, we plotted the orientation maps of where A1, A2 and A3 parameters were varied. We observed minimal changes to the orientation selectivity (Figure \ref{fig:para_A1}, \ref{fig:para_A2}, \ref{fig:para_A3}).

To ensure that the simulated stimulus was able to elicit a stimulus-consistent pattern among lamina, which is convinced to be a direct downstream of the photoreceptors, we examined the spatial and temporal activation patterns of lamina cells L1-L3 and their downstream neurons. Since our stimuli (OFF-bars) presented a straight bar pattern, we examined the firing rate of lamina cells when the OFF-bars were at different distances from them. We investigated the firing rate of lamina cells in response to the OFF-bars positioned at varying distances from them. Our findings revealed a clear trend: as the distance between the OFF-bars and the cell increased, the firing rate decreased correspondingly. Notably, the lamina cells (L1-L3) exhibited a peak firing rate when the OFF-bars were precisely aligned with the cell’s location (Figure \ref{fig:ex_fig1b}). These results confirmed that artificial stimulus-driven lamina neuron activity in our digital \textit{Drosophila} brain was consistent with the L1-L3 response properties known from biological experiments \cite{Ketkar_Burak}. 

We then sampled a set of downstream neurons - specifically Tm1, Tm2, Tm4, Mi1, Mi4 and Tm3, which are known from anatomical studies to be postsynaptic targets of L1-L3 neurons \cite{Mastsliah_Yu}. These neuron types were not shown in Figure \ref{fig:fig1b} to maintain visual simplicity, but their connectivity is well established. We randomly selected a single neuron from each of the Tm1, Tm2 and Tm4 subtypes as they are known OFF-cells \cite{Mastsliah_Yu} and evaluated if they exhibited firing patterns that reassemble the expected response profile to OFF-bar stimuli - that is, peaking when the bar aligns with their receptive field and decreasing with distance. As shown in Figure \ref{fig:ex_fig1c}, these OFF-responsive downstream neurons exhibited patterns as expected. In contrast, we also sampled a single neuron from each of the Mi1, Mi4 and Tm3 subtypes, known ON-cells \cite{Mastsliah_Yu}, and evaluated whether they showed the opposite trend - stronger inhibition, when the OFF-bar was nearby (Figure \ref{fig:ex_fig1d}). These neurons' spatial activation pattern mirrors the OFF-bar's location on the compound eye, consistent with known visual topology \cite{Mastsliah_Yu}. 


\subsection{Circular Gaussian function}
\label{sec:circular_gaussian_function}
We calculated the preferred orientations of each neuron using a circular Gaussian function defined as:
\begin{equation}
    \text{FiringRate}_n(\theta) = C \cdot \exp(-\frac{d(\theta, A)^2}{B}) + D,
\end{equation}
where:
\begin{itemize}
    \item \( \text{FiringRate}_n(\theta) \): predicted firing rate of neuron \(n\) at the orientation \(\theta\)
    \item \( \theta \): input orientation (in degrees, $0\degree \le \theta < 180\degree$)
    \item \( A \): preferred orientation (in degrees, fit parameters)
    \item \( B \): width parameters (sharpness of tuning)
    \item \( C \): amplitude of the tuning curve
    \item \( D \): baseline firing rate
    \item \( d(\theta, A) = \min(|\theta-A|, 180 - |\theta - A|) \): smallest circular distance between input orientation and preferred orientation.
\end{itemize}

\subsection{Circular standard deviation}
\label{sec:circular_standard_deviation}
We quantified the sharpness of orientation tuning within each column using the circular standard deviation (see Figure \ref{fig:fig5d}) in the \texttt{scipy} Python library, defined as:
\begin{equation}
    s = \sqrt{-2 \ln R}
\end{equation}
where:
\begin{itemize}
    \item \(s\): circular standard deviation (in radians)
    \item \(R\): mean resultant length, computed as:
    \begin{equation}
        R = \frac{1}{n}\sqrt{(\sum^{n}_{i=1} \cos{\alpha_i})^2 + (\sum^{n}_{i=1} \sin{\alpha_i})^2}
    \end{equation}
    \item \(\alpha_i\): preferred orientation of neuron \(i\) (in radians)
    \item \(n\): number of neurons in the column
\end{itemize}
This metric captured how tightly clustered the orientations are around the mean. A value of \(s=0\) indicates perfect alignment, with larger values reflecting broader tuning distributions.

\section{Targeted ablation analysis}
\label{sec:ablation}
To evaluate the contribution of specific upstream visual pathways to orientation selectivity, we performed targeted silencing of key input neuron populations. Using the equation: $Z\% =X / Y$, where $X$ represents the number of good fits in ablation, $Y$ represents the number of good fits in the original result, and $Z$ is the percentage of neurons remaining that still exhibit orientation selectivity. Silencing Tm neurons resulted in an 89.6\% reduction in well-fit orientation-selective neurons in the Pm layer but only a 9.3\% reduction in Dm. In contrast, silencing Mi neurons eliminated tuning across both layers, leaving fewer than 5\% of well-fit neurons in either Dm or Pm (Table~\ref{tab:ablation_summary}). These findings suggest that Mi neurons represent essential upstream sources of orientation selectivity throughout the optic lobe.

\begin{table}
  \caption{Ablation study results.}
  \label{tab:ablation_summary}
  \centering
  \begin{tabular}{lll}
    \toprule
    Z        & Tm silenced     & Mi silenced   \\
    \midrule
    Dm       & 90.7\%          & 2.4\%         \\
    Pm       & 10.4\%          & 3.9\%         \\
    \bottomrule
  \end{tabular}
\end{table}

\section{Orientation-based synaptic connectivity structure}
\label{sec:orientation_connectivity}
\begin{figure}[tp]
    \centering
    \begin{subfigure}[b]{0.49\linewidth}
        \centering
        \includegraphics[width=\textwidth]{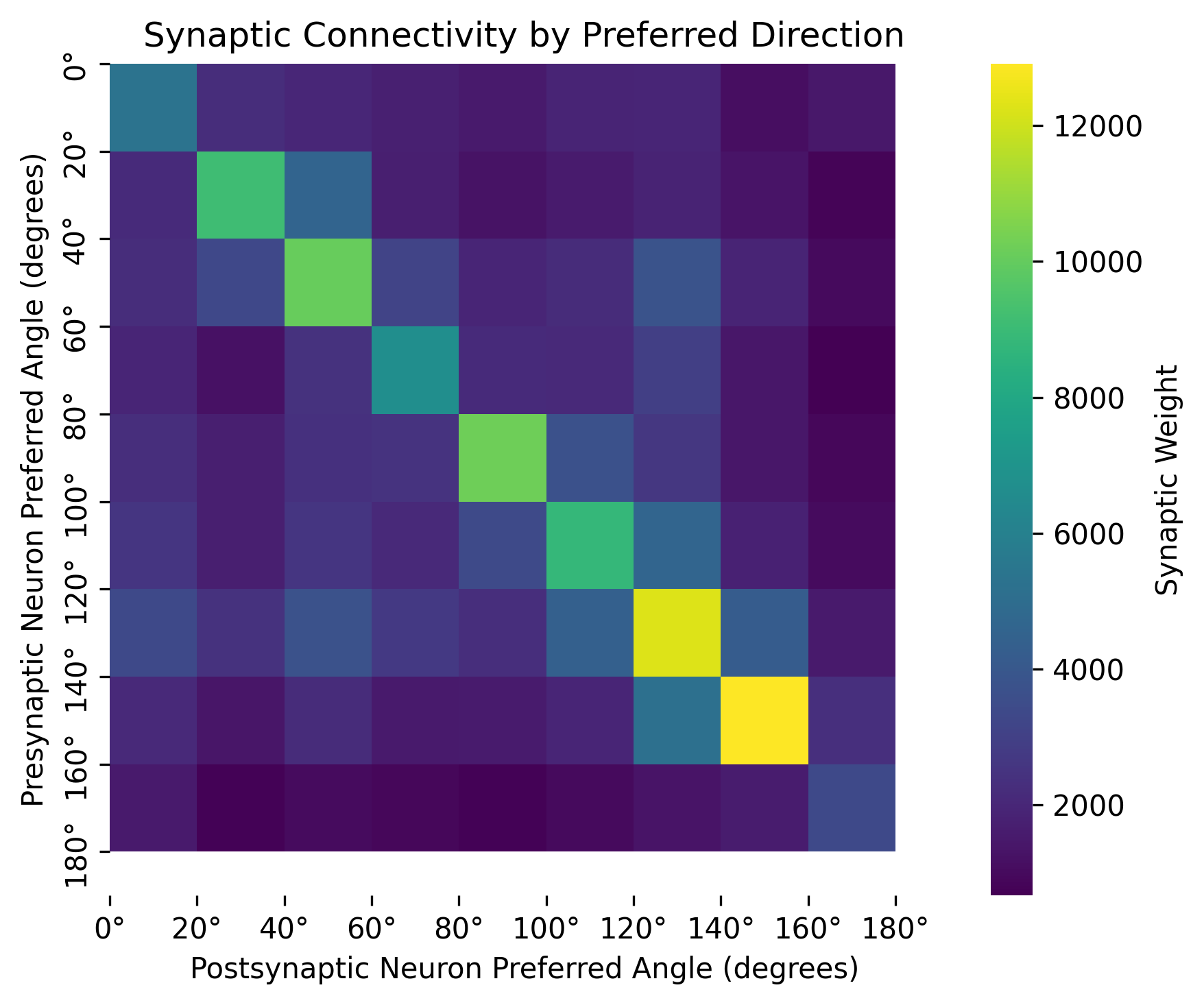}
        \caption{}
        \label{fig:con_exc}
    \end{subfigure}
    \begin{subfigure}[b]{0.49\linewidth}
        \centering
        \includegraphics[width=\textwidth]{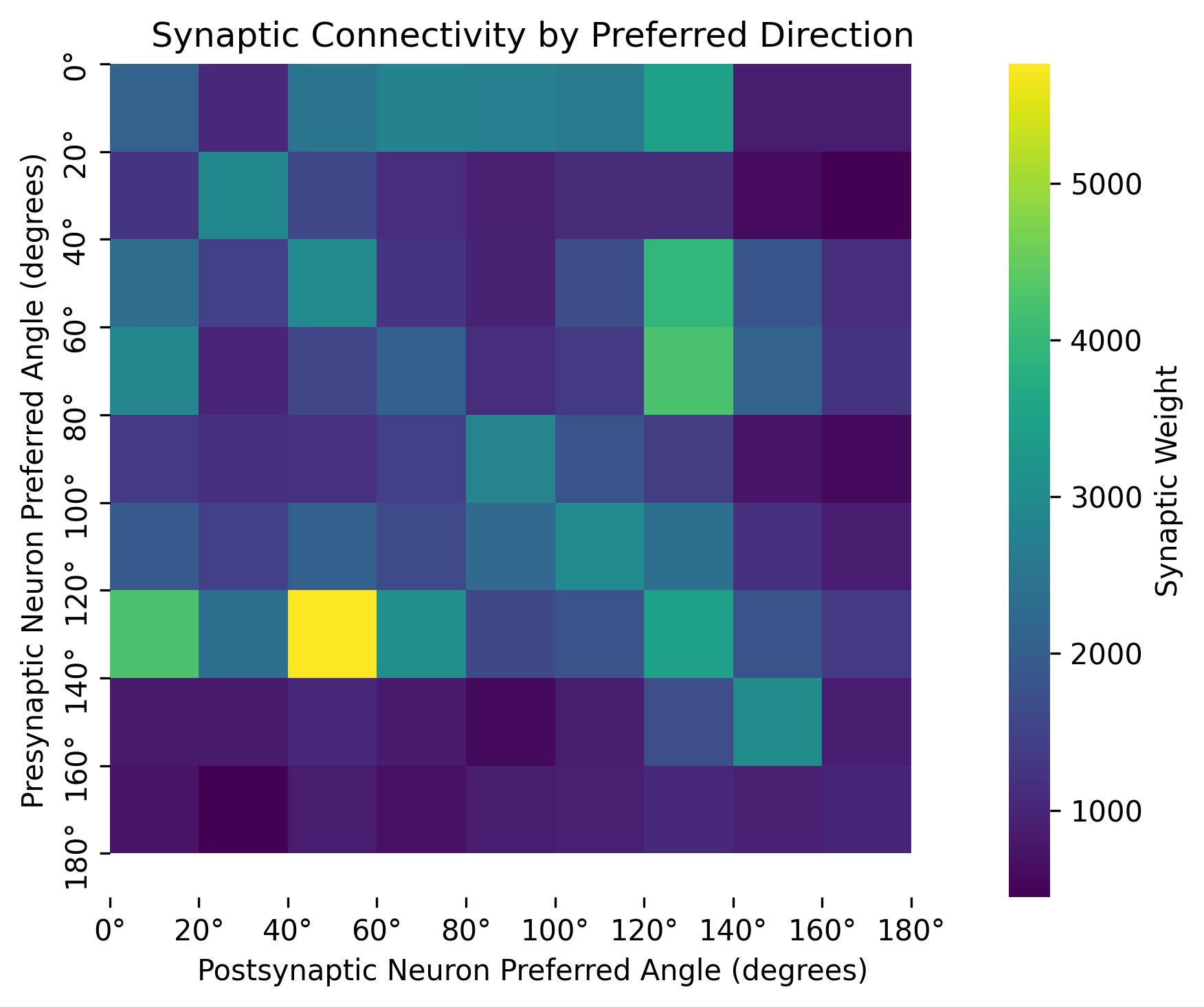}
        \caption{}
        \label{fig:con_inh}
    \end{subfigure}
    \caption{\textbf{Orientation-specific synaptic connectivity.} (a) Excitatory network showing strong diagonal structure, indicating neurons with similar orientation preferences preferentially connect. (b) Inhibitory network showing off-diagonal structure, consistent with cross-orientation suppression motifs.}
    \label{fig:connectivity}
\end{figure}

To examine whether recurrent connections in the connectome reflect functional similarity, we quantified synaptic connectivity as a function of orientation preference. Each synapse $(p_i, q_i)$ was represented as a point $(\theta_{p_i}, \theta_{q_i})$, where $\theta$ denotes the preferred orientation (in degrees) of the pre- and postsynaptic neurons, respectively. In the excitatory subnetwork, connections cluster along the diagonal ($\theta_{p_i} \approx \theta_{q_i}$), indicating that neurons with similar orientation preferences are more likely to be recurrently connected (Figure \ref{fig:con_exc}). This pattern is consistent with topographic organisation and colinear facilitation as proposed by \citet{Seung}, supporting the notion of like-to-like connectivity in local circuits. In contrast, inhibitory connections exhibit off-diagonal structure, suggesting cross-orientation interactions and selective suppression mechanisms (Figure \ref{fig:con_inh}). These results suggest that recurrent circuits in the \textit{Drosophila} optic lobe preferentially link neurons with similar selectivity, providing network-level evidence for structured functional organisation within the connectome.

\section{Extended figures}
\begin{figure}[tp]
    \centering
    \begin{subfigure}[b]{\linewidth} 
        \centering
        \includegraphics[width=\textwidth]{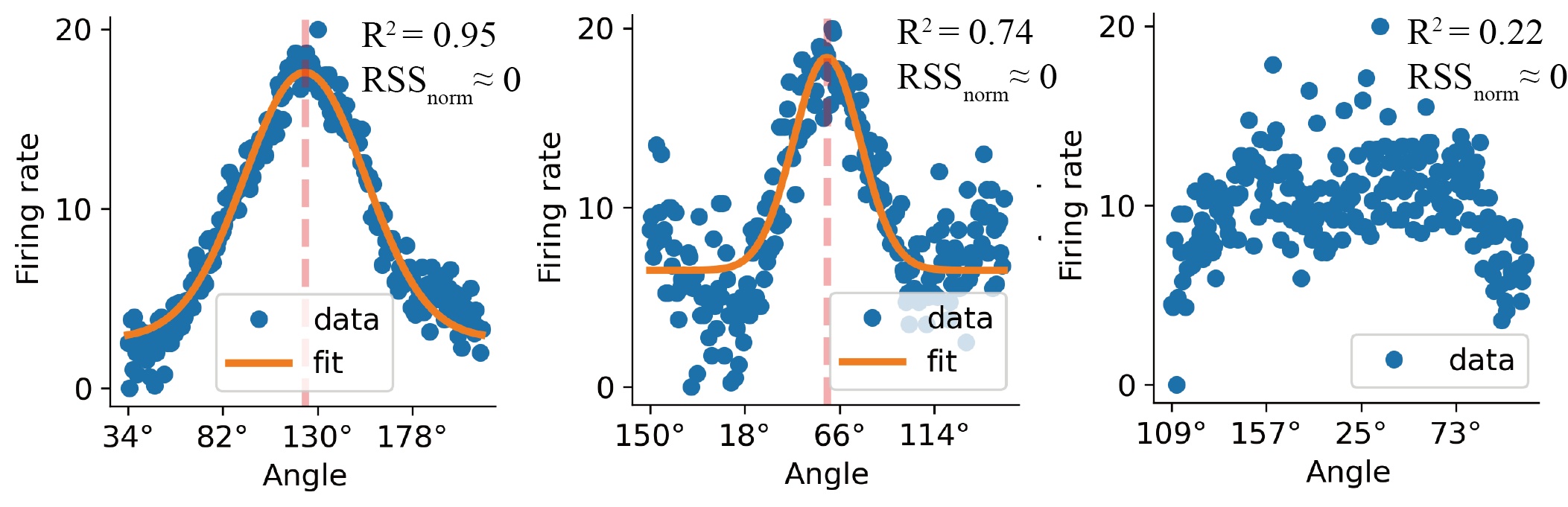}
        \caption{}
        \label{fig:gaussian_1a}
    \end{subfigure}
    \hfill
    \begin{subfigure}[b]{0.45\linewidth}
        \centering
        \includegraphics[width=\textwidth]{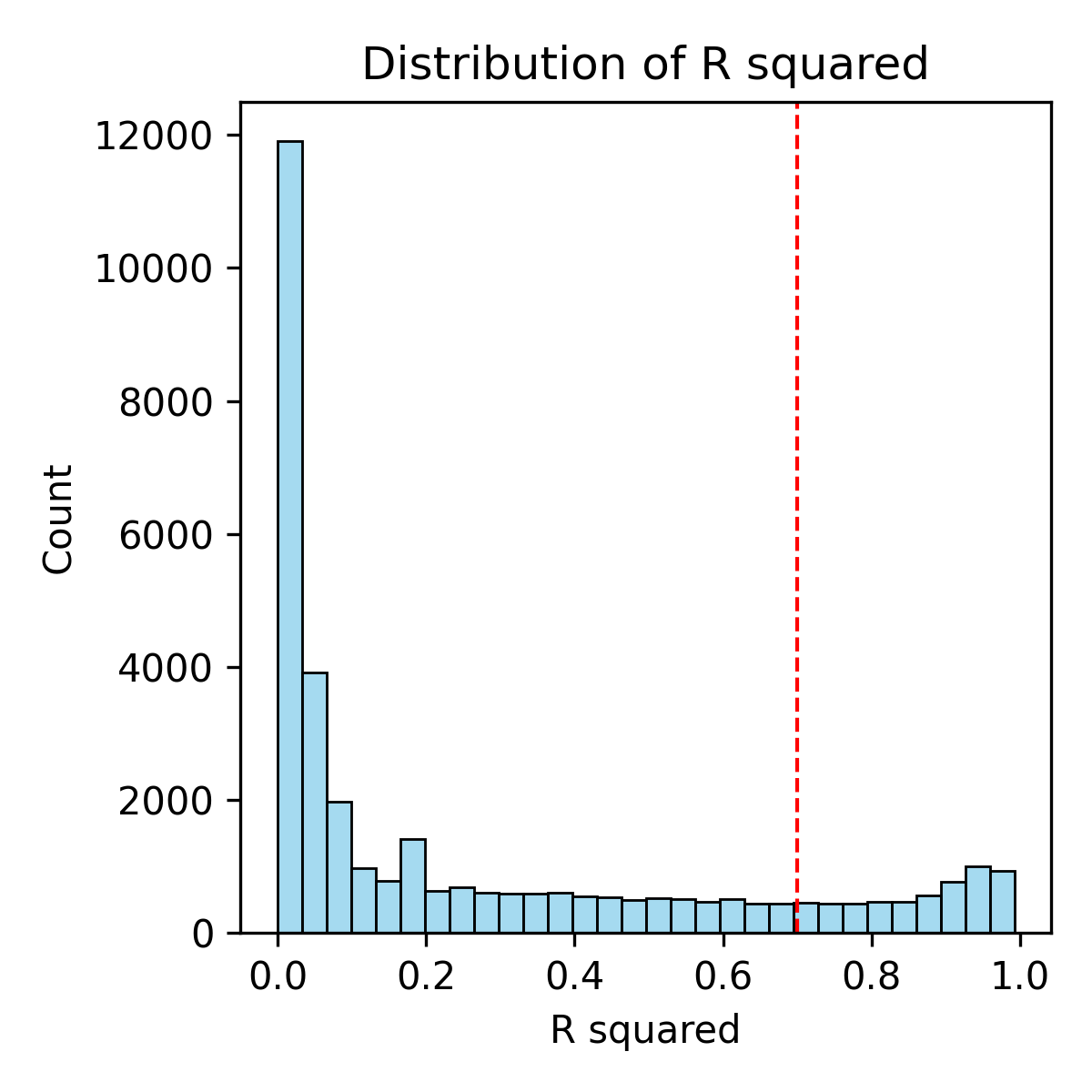}
        \caption{}
        \label{fig:gaussian_1b}
    \end{subfigure}
    \hfill
    \begin{subfigure}[b]{0.45\linewidth}
        \centering
        \includegraphics[width=\textwidth]{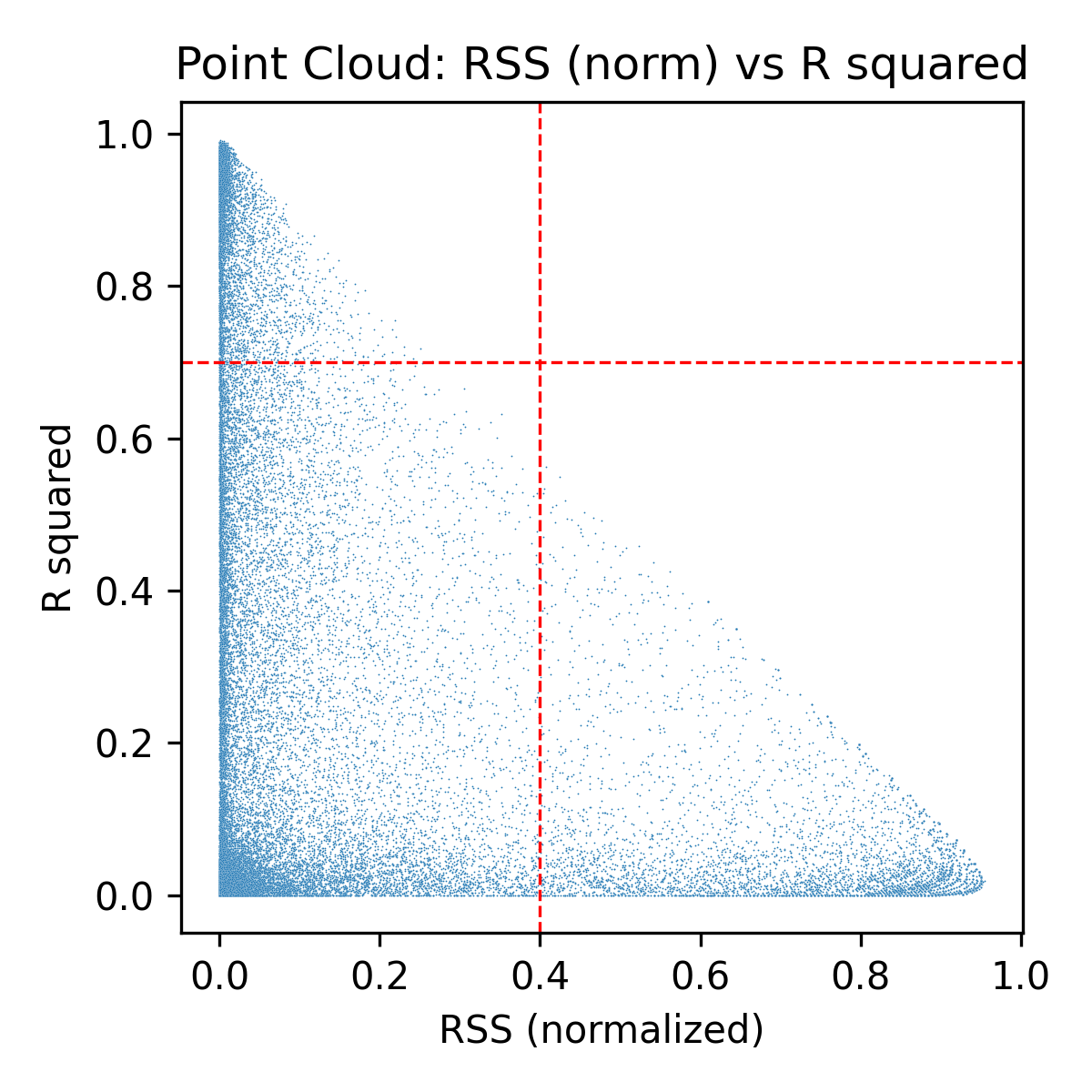}
        \caption{}
        \label{fig:gaussian_1c}
    \end{subfigure}
    
    \caption{\textbf{Model Gaussian fit quality.} (a) Representative neurons where their Gaussian fit is considered good (left), near borderline (middle) and poor (right). (b) Distribution of $R^2$ values, where the red line indicates the threshold. (c) Distribution of $R^2$ and normalised $RSS_{norm}$ of all neurons in the right optic lobe, including neurons in Figure \ref{fig:gaussian_1a}.}
    \label{fig:gaussian}
\end{figure}

\begin{figure}[tp]
    \centering
    \begin{subfigure}[b]{0.49\linewidth}
        \centering
        \includegraphics[width=\textwidth]{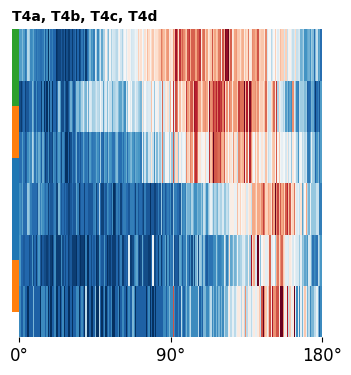}
        \caption{}
        \label{fig:ex_fig2a}
    \end{subfigure}
    \begin{subfigure}[b]{0.49\linewidth}
        \centering
        \includegraphics[width=\textwidth]{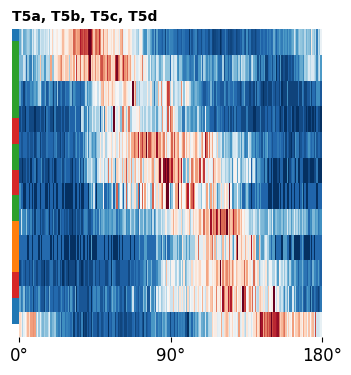}
        \caption{}
        \label{fig:ex_fig2b}
    \end{subfigure}
    \caption{\textbf{Orientation tuning profiles of some T4 and T5 neurons.} (a) Heatmap showing the orientation tuning responses of T4 neurons that are well-fitted. (b) Heatmap showing the orientation tuning responses of T5 that are well-fitted.}
    \label{fig:ex_fig2}
\end{figure}

\begin{figure}[tp]
    \centering
    \includegraphics[width=\textwidth]{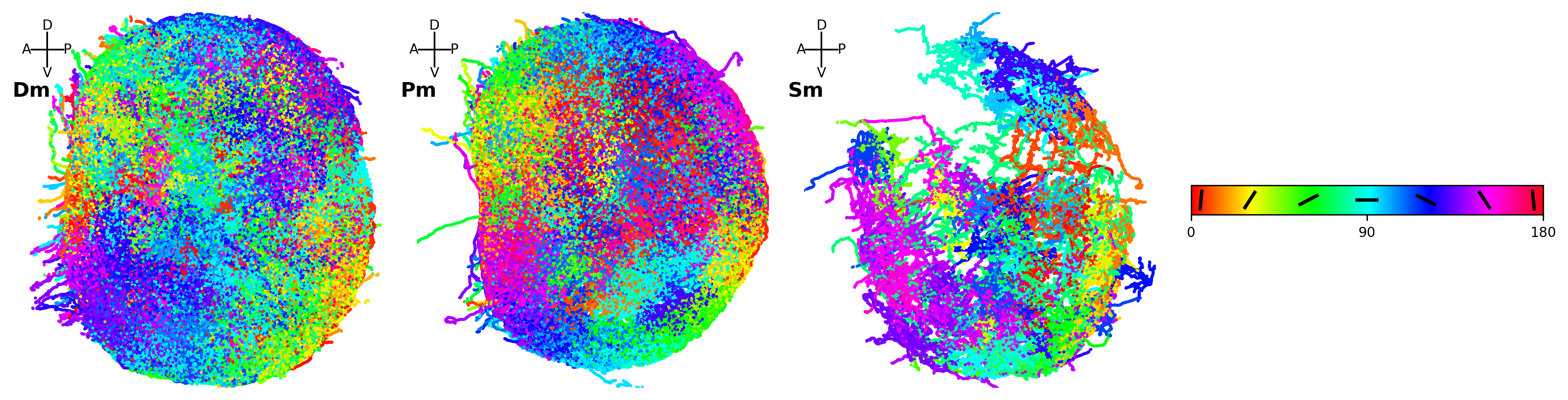}
    \caption{\textbf{Orientation preference structure across different neuron types.} Preferred orientation maps across optic lobe layers before smoothing, colour-coded by angle.} 
    \label{fig:ex_fig3}
\end{figure}

Figure \ref{fig:gaussian} shows the Gaussian model fitting used to quantify orientation selectivity. Figure \ref{fig:ex_fig2} shows the orientation tuning profile of T4 and T5 subtypes. Figure \ref{fig:ex_fig3} presents additional visualisation of orientation maps across different neuron types and layers within the optic lobe.

\end{document}